\documentclass[preprint,12pt]{aastex}

\def\msun{{\rm\,M_\odot}}
\newcommand{\etal}{et al.\ }

\slugcomment{Submitted to ApJ}

\begin{document}

\title{Properties of Cold Dark Matter Halos at $z>6$}

\author{Renyue Cen\altaffilmark{1}, Feng Dong\altaffilmark{2}, Paul Bode\altaffilmark{3}, and Jeremiah P. Ostriker\altaffilmark{4}}

\altaffiltext{1} {Princeton University Observatory, 
Princeton University, Princeton, NJ 08544; cen@astro.princeton.edu}

\altaffiltext{1} {Princeton University Observatory, 
Princeton University, Princeton, NJ 08544; feng@astro.princeton.edu}

\altaffiltext{1} {Princeton University Observatory, 
Princeton University, Princeton, NJ 08544; bode@astro.princeton.edu}

\altaffiltext{1} {Princeton University Observatory, 
Princeton University, Princeton, NJ 08544; jpo@astro.princeton.edu;
Institute for Astronomy,
Cambridge University, Cambridge, England; jpo@ast.cam.ac.uk}

\accepted{ }

\begin{abstract}
We compute the properties of dark matter halos 
with mass $10^{6.5}-10^9\msun$
at redshift $z=6-11$ in the standard cold dark matter
cosmological model, utilizing a very high resolution
N-body simulation.
We find that dark matter halos in these mass and redshift ranges 
are significantly biased
over matter with a bias factor in the range $2-6$.
The dark matter halo mass function displays a 
slope of $2.05\pm 0.15$ at the small mass end.
We do not find a universal dark matter density profile.
Instead, we find
a significant 
dependence of the central density 
profile of dark matter halos on halo mass and epoch
with $\alpha_0=0.4-1.0$;
the high-mass ($M\ge 10^8\msun$)
low-redshift ($z\sim 6$) halos occupy the high end
of the range and low-mass ($M\sim 10^{7}\msun$)
high-redshift ($z\sim 11$) halos occupy the low end.
Additionally, for fixed mass and epoch there is 
a significant dispersion in $\alpha_0$ 
due to the stochastic assembly of halos.
Our results fit a relationship of the form 
$\alpha_0=0.75((1+z)/7.0)^{-1.25}(M/10^7\msun)^{0.11(1+z)/7.0}$
with a dispersion about this fit of $\pm 0.5$ and no systematic
dependence of variance correlated with environment.
The median spin parameter of dark matter halos
is $0.03-0.04$ but with a large lognormal dispersion of $\sim 0.4$.
Various quantities are tabulated or fitted with empirical formulae.

\end{abstract}


\section{Introduction}

The reionization epoch is now within the direct
observational reach thanks to rapid recent observational advances 
in two fronts ---
optical quasar absorption from Sloan Digital Sky Survey (SDSS)
(Fan \etal 2001; Becker \etal 2001) 
and the Wilkinson Microwave Anisotropy Probe (WMAP) experiment
(Kogut \etal 2003).
The picture painted by the combined observations, 
perhaps not too surprisingly,
strongly suggests
a complex cosmological reionization process,
consistent with the double reionization scenario (Cen 2003). 
It may be that this is the beginning 
of a paradigm shift in our focus on the high redshift universe:
the star formation history of the early universe
can now be observationally constrained.

It thus becomes urgent to theoretically explore
galaxy and star formation process 
at high redshift in the dark age ($z\ge 6$).
In the context of the standard cold dark matter model
it is expected that stars within halos 
of mass $10^7-10^9\msun$ at high redshift
play an important, if not dominant, role
in determining how and when the universe was reionized. 
Furthermore, these fossil halos may be seen
in the local universe as satellites of giant galaxies.
This linkage may potentially provide 
a great leverage to nail down the properties of the 
high redshift galaxies.

In this paper, as a step towards understanding
galaxy formation at high redshift,
we investigate the properties of dark matter halos at $z\ge 6$,  
using very high resolution TPM N-body (Bode \etal 2001; Bode \& Ostriker 2003)
simulations.
While there is an extensive literature 
on properties of halos at low redshift,
there is virtually no systematic study of dark halos
at $z\ge 6$.  
The LCDM simulation has a comoving box size of $4h^{-1}$Mpc 
with $512^3=10^{8.2}$ particles,
a particle mass of $m_p=3.6\times 10^4 \ h^{-1}\msun$,
and comoving gravitational softening length of $0.14 \ h^{-1}$kpc.
These resolutions allow us to accurately characterize
the properties of halos 
down to a mass $10^{6.5} \ h^{-1}\msun$ (having about $100$ particles
within the virial radius).
The outline of this paper is as follows.
The simulation details are given in \S 2.
In \S 3 we quantify properties of dark matter halos in the mass range
$10^{6.5}-10^9\msun$, including the mass function, 
bias and clustering properties,
density profile distribution, angular momentum spin 
parameter distribution,
internal angular momentum distribution
and peculiar velocity distribution.
We conclude in \S 4.

\section{The Simulation} \label{sec:thesim}

A standard spatially flat LCDM cosmology was chosen,
with $\Omega_m=0.27$ and $\Omega_\Lambda=0.73$; the
Hubble constant was taken to be 70 km/s/Mpc.
The initial conditions were created using the GRAFIC2
package by Bertschinger (2001). 
The matter transfer 
function was calculated with the included Boltzmann integrator
(Ma \& Bertschinger 1995),
using $\Omega_b h^2=0.211$ for the baryon
fraction and $\sigma_8=0.73$ for the normalization of the matter 
power spectrum.

The simulation contained $N=512^3$ particles in a comoving
periodic box 4$h^{-1}$Mpc on a side, making the particle mass
$m_p=3.57\times 10^4 \ h^{-1}M_\odot$.  The starting redshift
was $z=53$, and the system was evolved down to $z=6$.
The evolution was carried out with the parallel
Tree--Particle--Mesh code TPM (Xu 1995; Bode, Ostriker, \& Xu 2000;
Bode \& Ostriker 2003),
using a $1024^3$ mesh.  The evolution took
1150 PM steps, with particles in dense regions taking up to
19,500 steps.  The run was carried out using up to 256
processors on the Terascale Computing System at Pittsburgh
Supercomputing Center.

The spline softening length in the tree portion of the
code was set to $\epsilon=0.14 \ h^{-1}$kpc. 
With this softening length, relaxation by $z=6$ inside the core
of a collapsed halo (assuming an NFW density distribution
with $c=12$) will not be significant over the course of
the simulation for those objects containing more than 100 particles.
The opening angle in the Barnes-Hut criterion used by TPM
was $\theta=0.577$, and the time step parameter $\eta=0.3$;
also, the initial values for locating trees were $A=2.0$ and
$B=12.5$--- see Bode \& Ostriker (2003) for details.  
In the TPM code, not all regions are treated
at full resolution.  The limiting density (above which all
cells are put into trees for increased resolution) rises
with time.  By the end of this run, all cells containing more 
than 18 particles are still being followed at full resolution.
Thus this factor is not important if the analysis is limited
to halos with over 100 particles.

Dark matter halos are identified 
using DENMAX scheme (Bertschinger \& Gelb 1991),
smoothing the density field with a Gaussian length of $300h^{-1}$kpc.
In computing all quantities we include all particles
located inside the virial radius of a halo.

\section{Results}

\epsscale{1.0}
\begin{figure}
\plotone{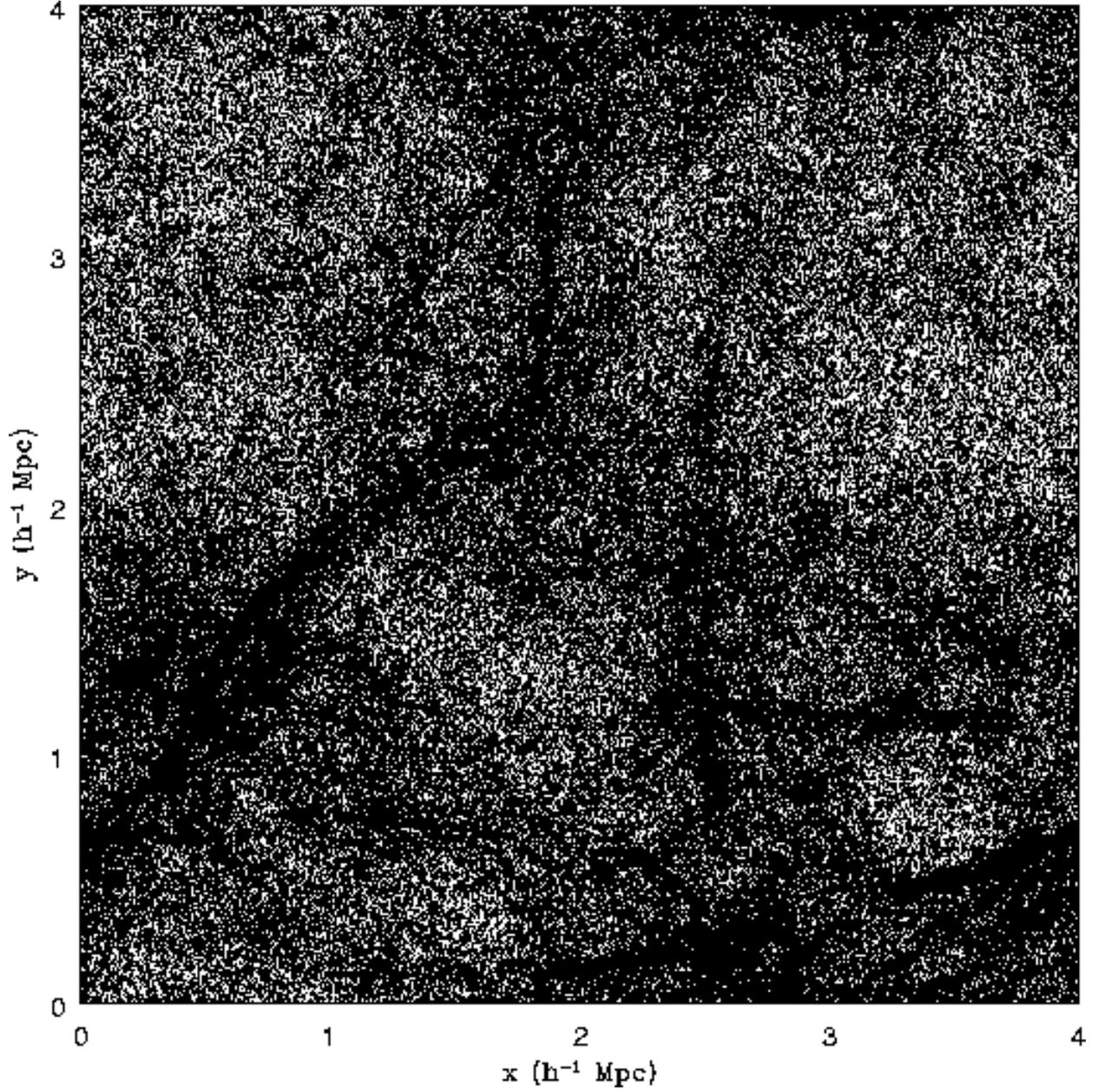}
\caption{The distribution of dark matter particles at $z=6$ 
projected onto the x-y plane (0.25$\%$ of the total).
\label{f1}}
\end{figure}

\begin{figure}
\centerline{\bf M $> 10^6 \ h^{-1} \msun$}
\vskip 0.01in
\includegraphics[width=1.0\hsize,angle=0]{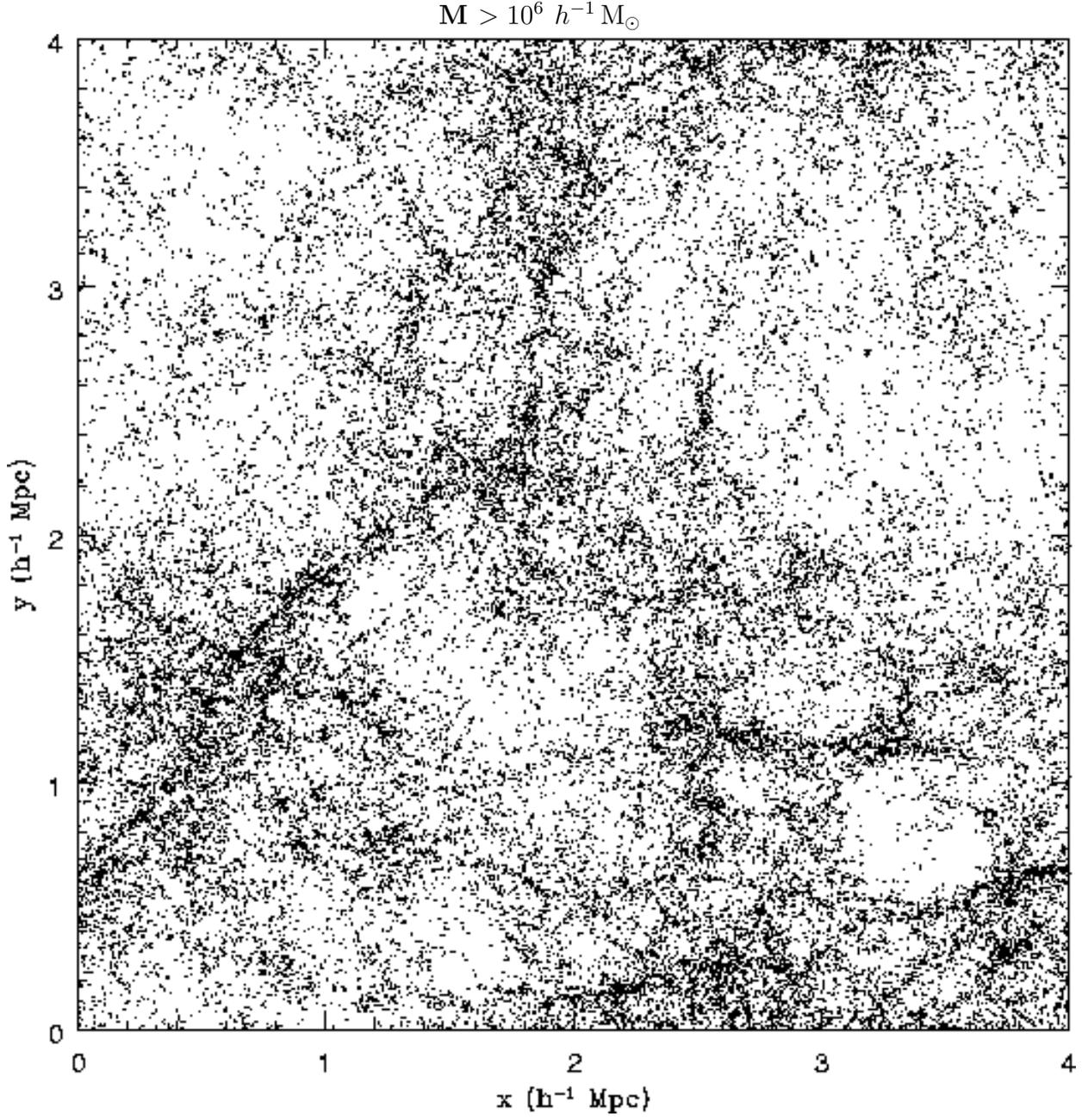}
\caption{The distributions of all dark matter halos 
with masses greater than $(10^6,10^{6.5},10^7,10^{7.5},10^8)\msun$, 
respectively, at $z=6$.
\label{f2a}}
\end{figure}
\begin{figure}
\centerline{\bf M $> 10^{6.5} \ h^{-1} \msun$}
\vskip 0.01in
\includegraphics[width=1.0\hsize,angle=0]{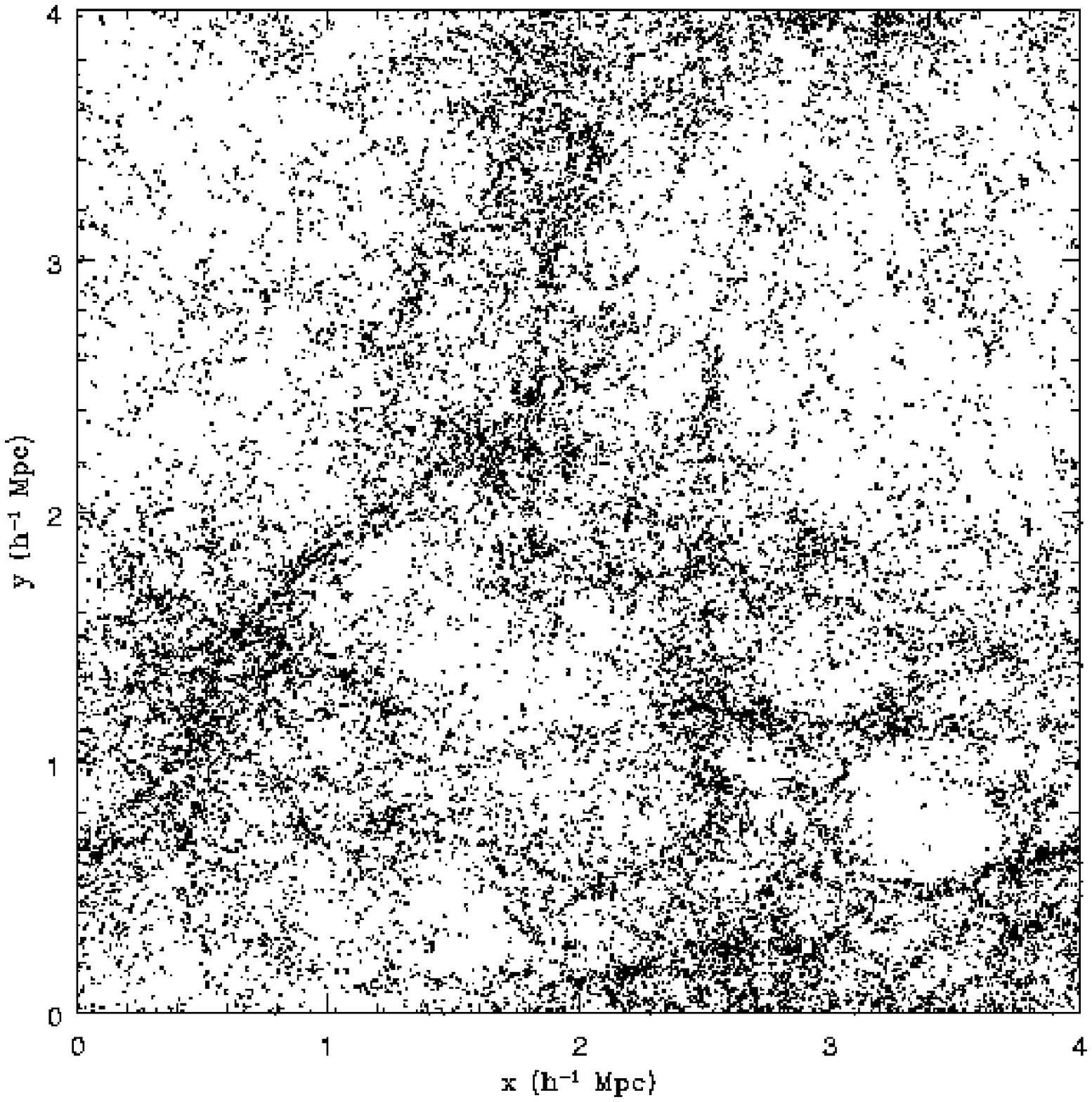}
\addtocounter{figure}{-1}
\caption{Continued.}
\end{figure}
\begin{figure}
\centerline{\bf M $> 10^6 \ h^{-1} \msun$}
\vskip 0.01in
\includegraphics[width=1.0\hsize,angle=0]{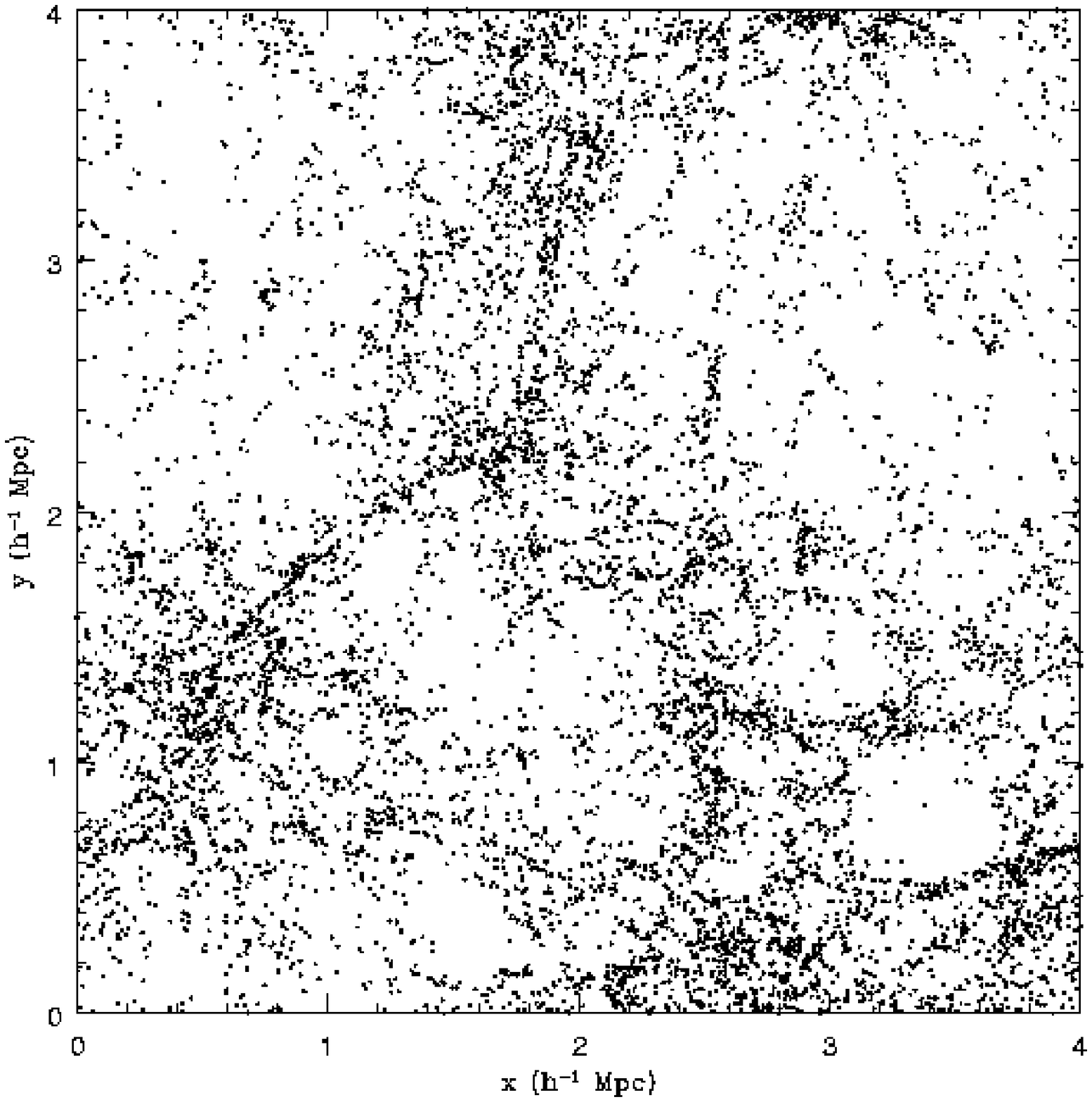}
\addtocounter{figure}{-1}
\caption{Continued.}
\end{figure}
\begin{figure}
\centerline{\bf M $> 10^{7.5} \ h^{-1} \msun$}
\vskip 0.01in
\includegraphics[width=1.0\hsize,angle=0]{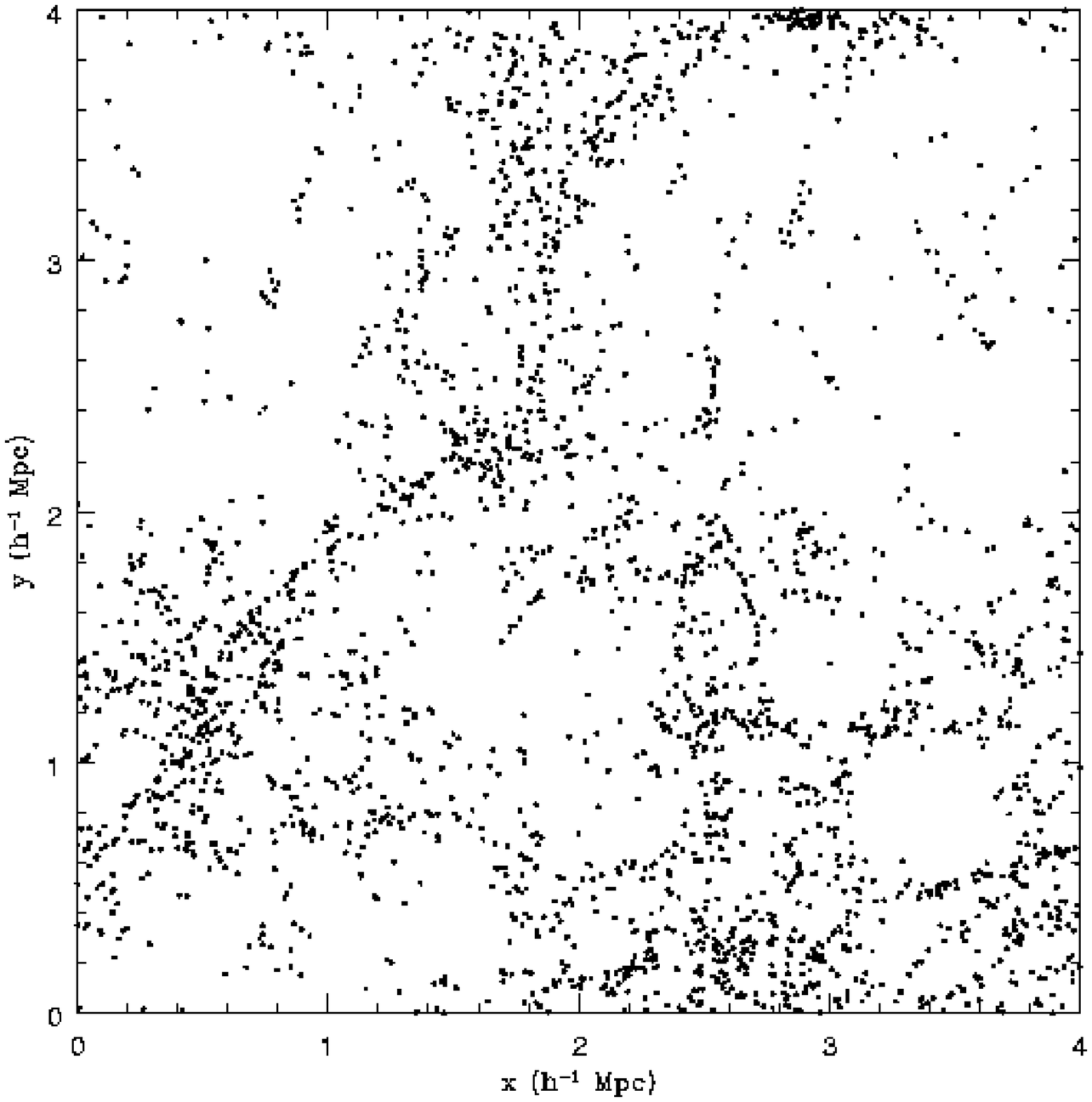}
\addtocounter{figure}{-1}
\caption{Continued.}
\end{figure}
\begin{figure}
\centerline{\bf M $> 10^8 \ h^{-1} \msun$}
\vskip 0.01in
\includegraphics[width=1.0\hsize,angle=0]{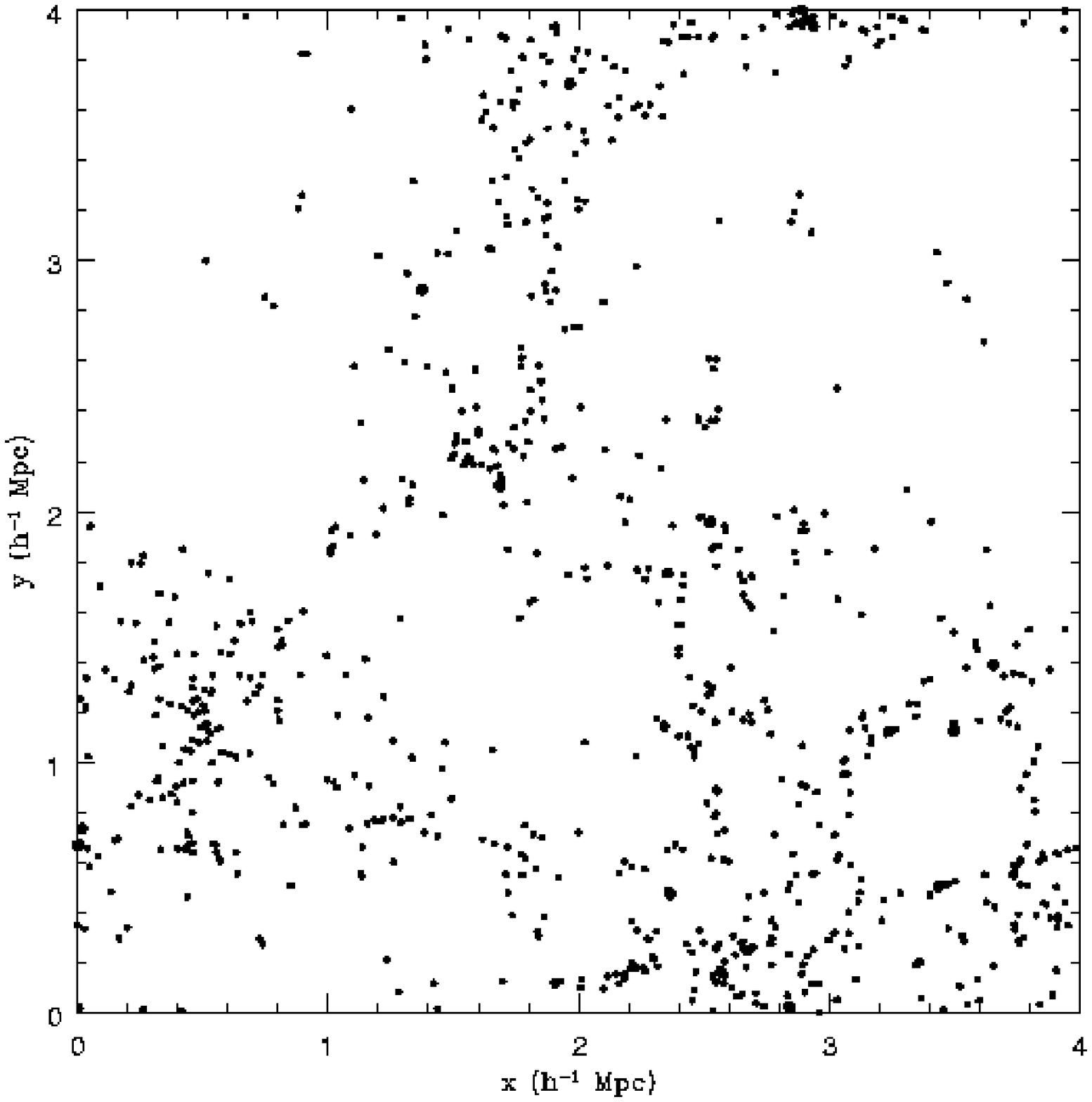}
\addtocounter{figure}{-1}
\caption{Continued.}
\end{figure}
\subsection{Pictures}
First, we present visually a distribution of the dark matter mass
and dark matter halos of varying masses.
Figure 1 shows the distribution of dark matter particles
projected onto the x-y plane.
Figures (2a,b,c,d,e) show 
the distributions of dark matter halos 
with masses greater than
$(10^6,10^{6.5},10^7,10^{7.5},10^8)\msun$, respectively, at $z=6$.
The progressively stronger clustering of more massive halos
is clearly visible in the display but we will return
to the clustering properties more quantitatively in \S 3.3.
It is also noted that voids are progressively more visible in
the higher mass halos than in low mass halos.

\subsection{Dark Matter Halo Mass Function}

\epsscale{0.8}
\begin{figure}
\plotone{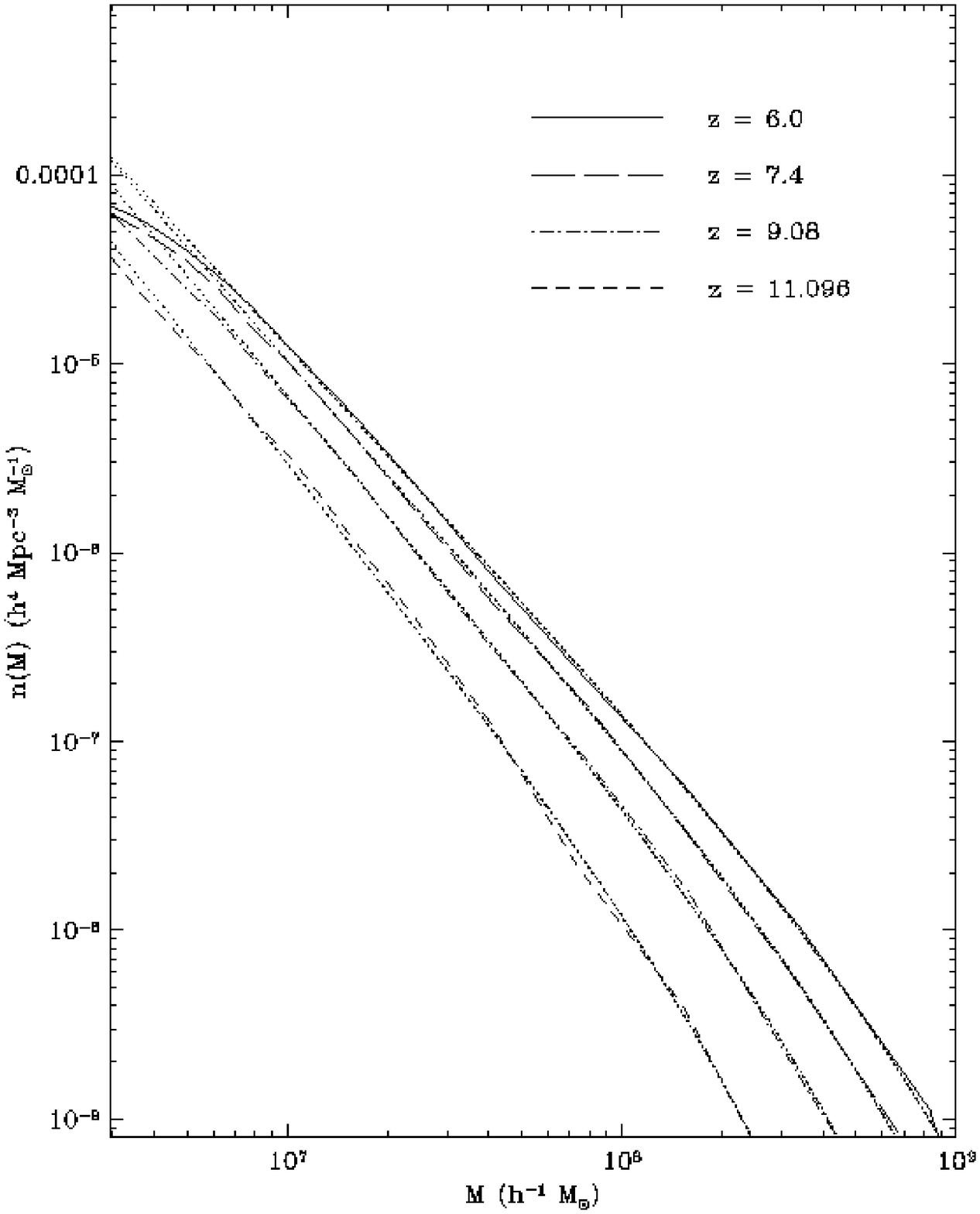}
\caption{The halo mass functions at four redshifts. Dotted lines represent the 
fitted Schechter functions with parameters summarized in Table 1. 
\label{f3}}
\end{figure}

\begin{deluxetable}{rrrrrrrr} 
\tablecolumns{5} 
\tablewidth{0pc} 
\tablecaption{Halo Mass Function Fitting Parameters} 
\tablehead{ 
\colhead{Parameters \ } & \colhead{\ \ \ \ z=6.0 \ }   & \colhead{\ \ z=7.4 \ }  & 
\colhead{z=9.08} & \colhead{\ \ z=11.096}}
\startdata 
$n_0$ ($h^3$ Mpc$^{-3}$) & 0.85 \ \ & 1.20 \ \ & 1.75 \ \ & 2.40 \ \  \\ 
\smallskip
$\alpha$ \ \ \ \ \ \ \ \ \  & 1.9 \ \ \ & 2.0 \ \ \ & 2.1 \ \ \ & 2.2 \ \ \ \\ 
\smallskip 
$M_*$ ($h^{-1}$ Mpc) & $8\times10^8$ & \ $6\times10^8$ & \ $4\times10^8$ & $2\times10^8$ \\ 
\hline
\enddata 
\end{deluxetable} 

Figure 3 shows the halo mass functions at four redshifts.
Table 1 summarizes the fitting parameters for a Schechter
function of the form 
\begin{equation}
n(M)dM = n_0 ({M\over M_*})^{-\alpha} \exp{(-M/M_*)} {dM\over M_*}.
\end{equation}
\noindent
We see that the Schechter function provides a good fit
to the computed halo mass function.
The faint end slope is 
$\alpha=2.05\pm 0.15$, consistent with the expectation
from Press-Schechter theory (Press \& Schechter 1974).
While there appears to be
a slight steepening of the slope at the low
mass end from $-1.9$ to $-2.2$ from $z=6$ to $z=11.1$,
it is unclear, however, how significant this trend is,
given the adopted, somewhat degenerate fitting formula.
The turnover at $M_h\sim 10^{6.5}\msun$
indicates the loss of validity of our simulation at the low mass end.

\subsection{Bias of Dark Matter Halos}

\begin{figure*}
\includegraphics[width=3.2in]{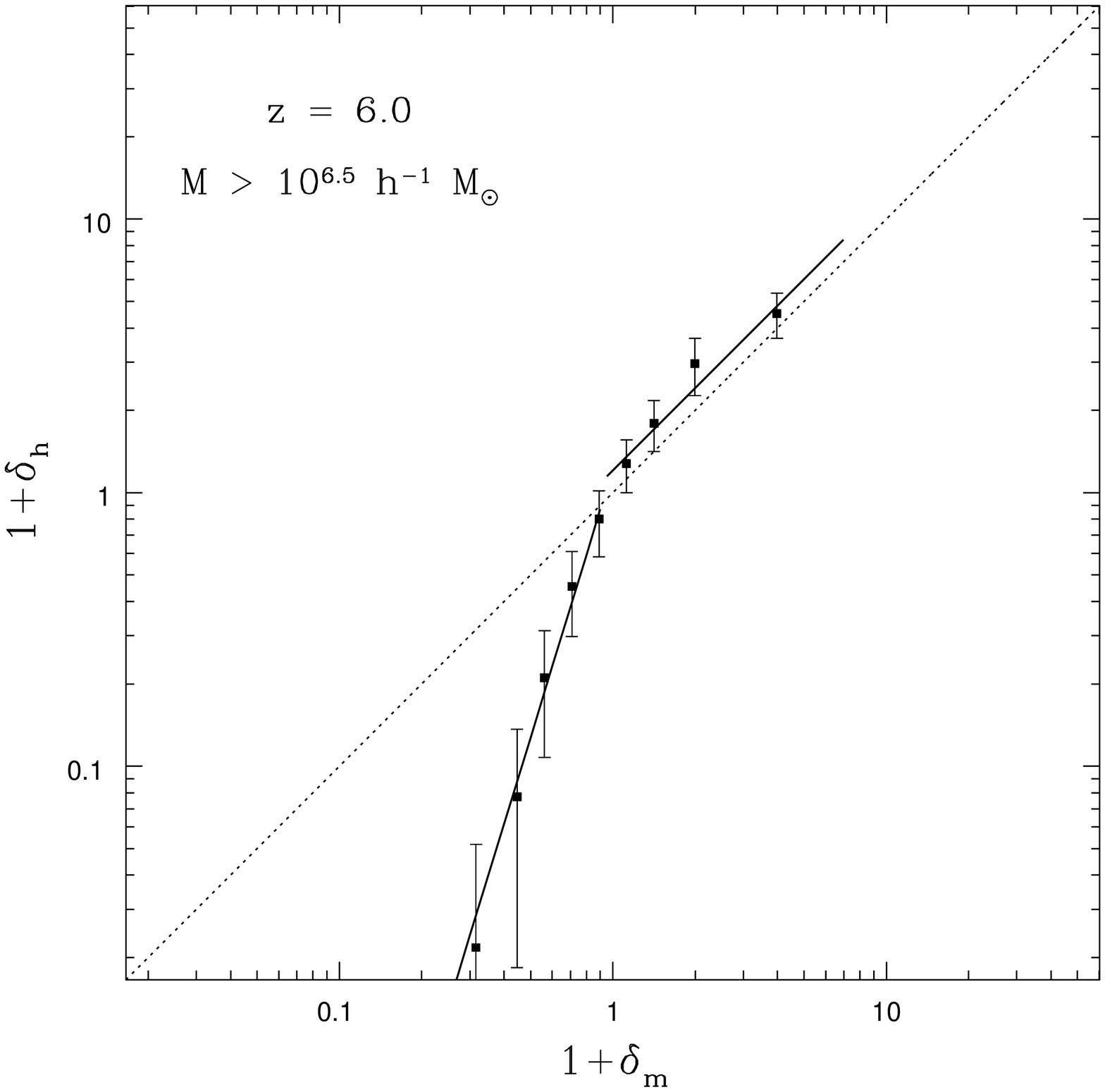}
\hspace{0.01in}
\includegraphics[width=3.2in]{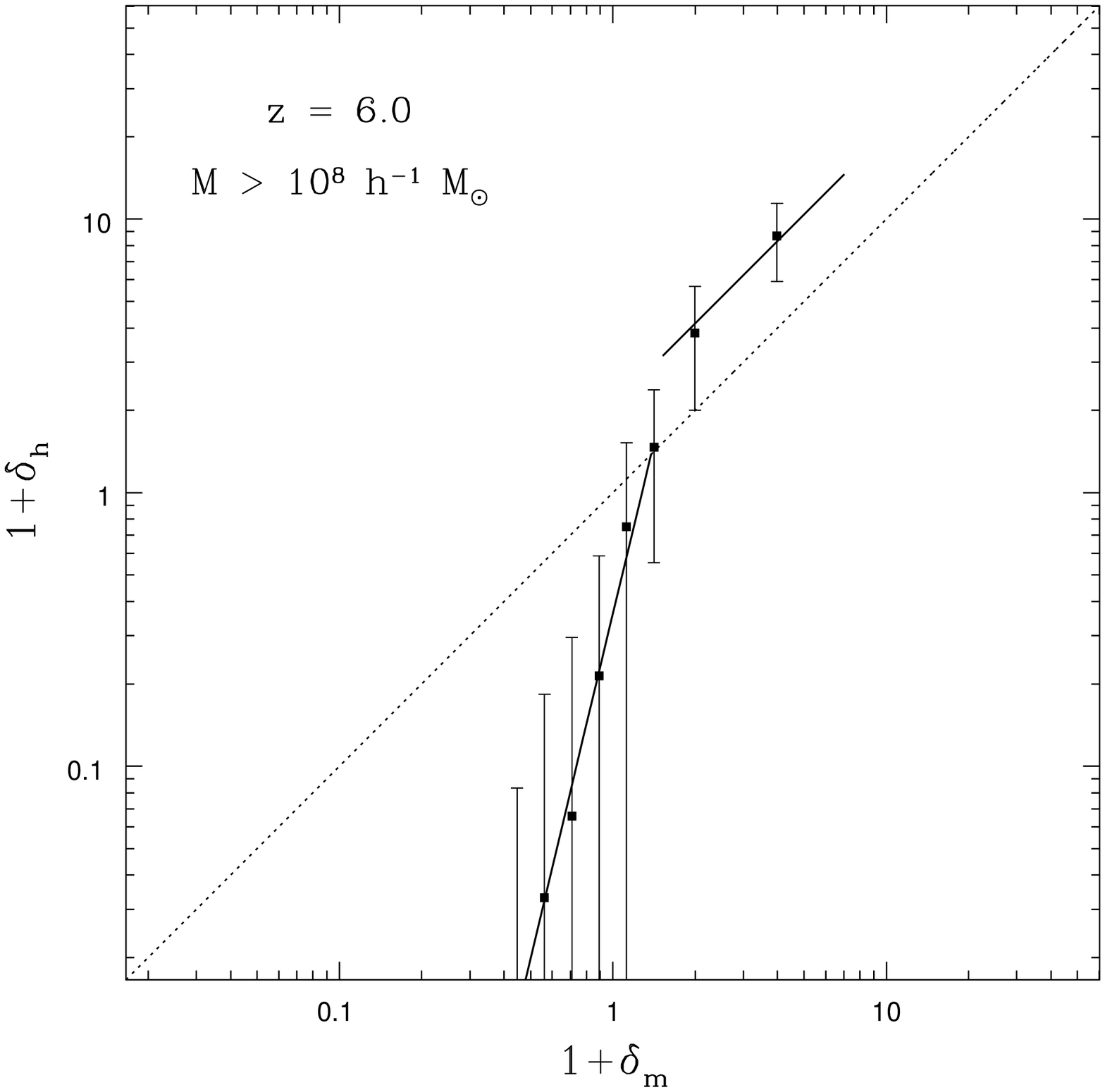}
\vskip 0.05in
\includegraphics[width=3.2in]{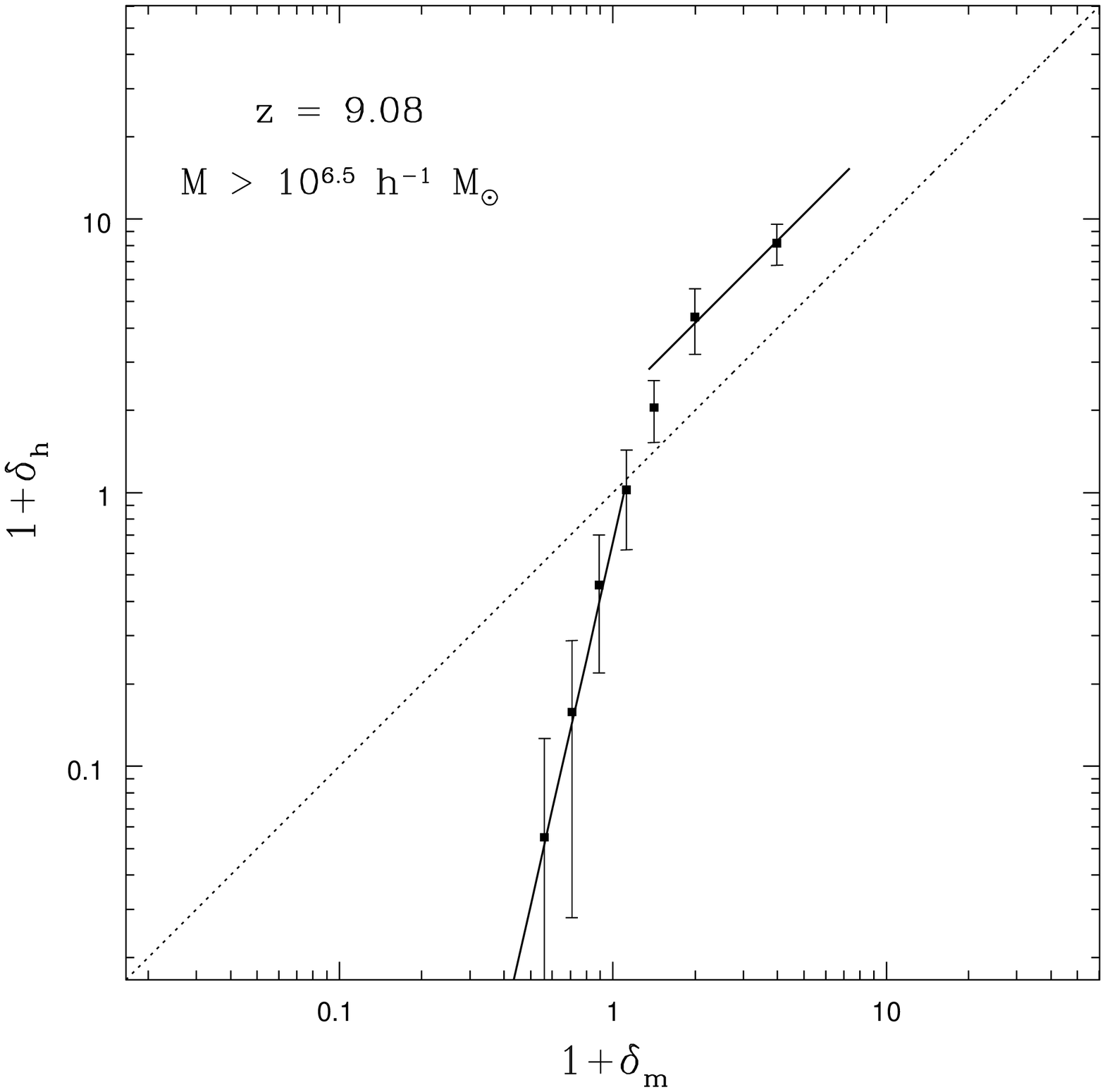}
\hspace{0.01in}
\includegraphics[width=3.2in]{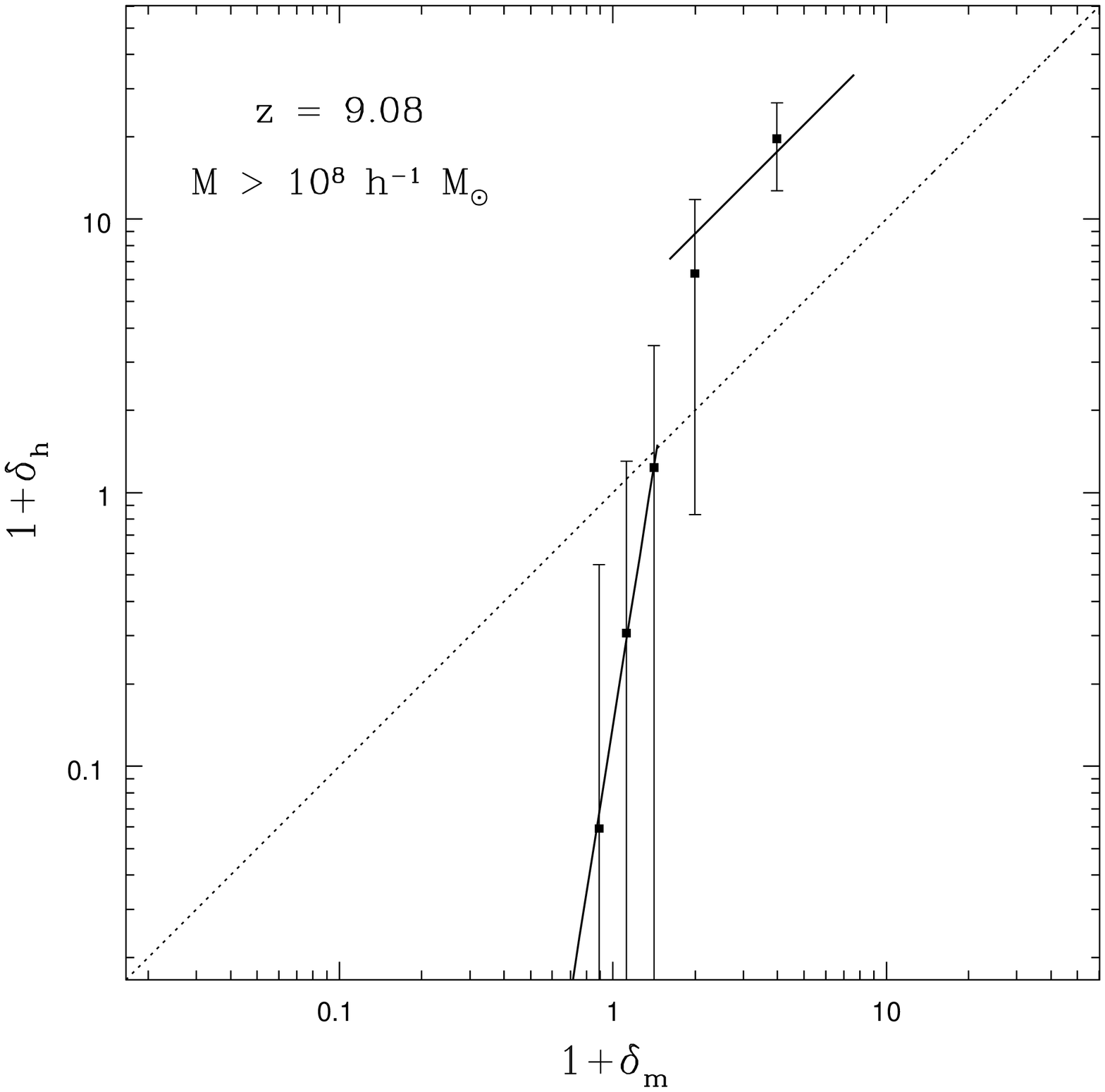}
\caption{Bias of dark matter halos in four randomly selected cases upon different 
mass and redshift.
The darkened lines show the fitting formulae of the bias relation. 
\label{f4}}
\end{figure*}

\begin{deluxetable}{rrrrrrrr} 
\tablecolumns{5} 
\tablewidth{0pc} 
\tablecaption{Halo Bias : b(M,z)} 
\tablehead{ 
\colhead{Halo Mass ($h^{-1}$ M$_\odot$) } & \colhead{\ \ \ \ z=6.0 \ }   & \colhead{\ \ z=7.4 \ }  & 
\colhead{z=9.08} & \colhead{\ \ z=11.096}}
\startdata 
$>$10$^{6.0}$ \ \ \ \ \ \ \ \ & \ 1.16$\pm$0.12 & \ \ \ 1.36$\pm$0.18 \ \ & \ 1.70$\pm$0.20 \ & 
\ \ 2.46$\pm$0.23 \\ 
$>$10$^{6.5}$ \ \ \ \ \ \ \ \ & \ 1.21$\pm$0.13 & \ \ \ 1.45$\pm$0.17 \ \ & \ 2.09$\pm$0.30 \ & 
\ \ 3.05$\pm$0.33 \\ 
$>$10$^{7.0}$ \ \ \ \ \ \ \ \ & \ 1.37$\pm$0.19 & \ \ \ 1.60$\pm$0.23 \ \ & \ 2.77$\pm$0.40 \ & 
\ \ 3.92$\pm$0.90 \\ 
$>$10$^{7.5}$ \ \ \ \ \ \ \ \ & \ 1.72$\pm$0.32 & \ \ \ 2.12$\pm$0.47 \ \ & \ 3.46$\pm$0.87 \ & 
\ \ 6.17$\pm$3.17 \\ 
$>$10$^{8.0}$ \ \ \ \ \ \ \ \ & \ 2.08$\pm$0.55 & \ \ \ 2.58$\pm$0.90 \ \ & \ 4.43$\pm$1.47 \ & 
\ \ 9.90$\pm$4.58 \\ 
\hline 
\enddata 
\end{deluxetable}

\begin{deluxetable}{rrrrrrrr} 
\tablecolumns{5} 
\tablewidth{0pc} 
\tablecaption{Halo Bias : c(M,z)} 
\tablehead{ 
\colhead{Halo Mass ($h^{-1}$ M$_\odot$)} & \colhead{\ \ \ \ z=6.0 \ }   & \colhead{\ \ z=7.4 \ }  & 
\colhead{z=9.08} & \colhead{\ \ z=11.096}}
\startdata 
$>$10$^{6.0}$ \ \ \ \ \ \ \ \ & \ 3.00$\pm$0.55 & \ \ \ 3.80$\pm$0.79 \ \ & \ 4.14$\pm$0.80 \ & 
\ \ 4.90$\pm$1.49 \\ 
$>$10$^{6.5}$ \ \ \ \ \ \ \ \ & \ 3.27$\pm$0.71 & \ \ \ 4.20$\pm$1.04 \ \ & \ 4.41$\pm$1.31 \ & 
\ \ 5.52$\pm$2.43 \\ 
$>$10$^{7.0}$ \ \ \ \ \ \ \ \ & \ 3.43$\pm$0.70 & \ \ \ 4.37$\pm$1.07 \ \ & \ 4.97$\pm$2.59 \ & 
\ \ 7.33$\pm$6.29 \\ 
$>$10$^{7.5}$ \ \ \ \ \ \ \ \ & \ 3.93$\pm$1.70 & \ \ \ 4.98$\pm$2.96 \ \ & \ 5.09$\pm$3.31 \ & 
\ \ 7.41$\pm$9.09 \\ 
$>$10$^{8.0}$ \ \ \ \ \ \ \ \ & \ 4.18$\pm$2.28 & \ \ \ 5.61$\pm$4.68 \ \ & \ 6.32$\pm$6.12 \ & 
\ \ 6.77$\pm$8.90 \\ 
\hline
\enddata 
\end{deluxetable} 

\epsscale{0.8}
\begin{figure}
\plotone{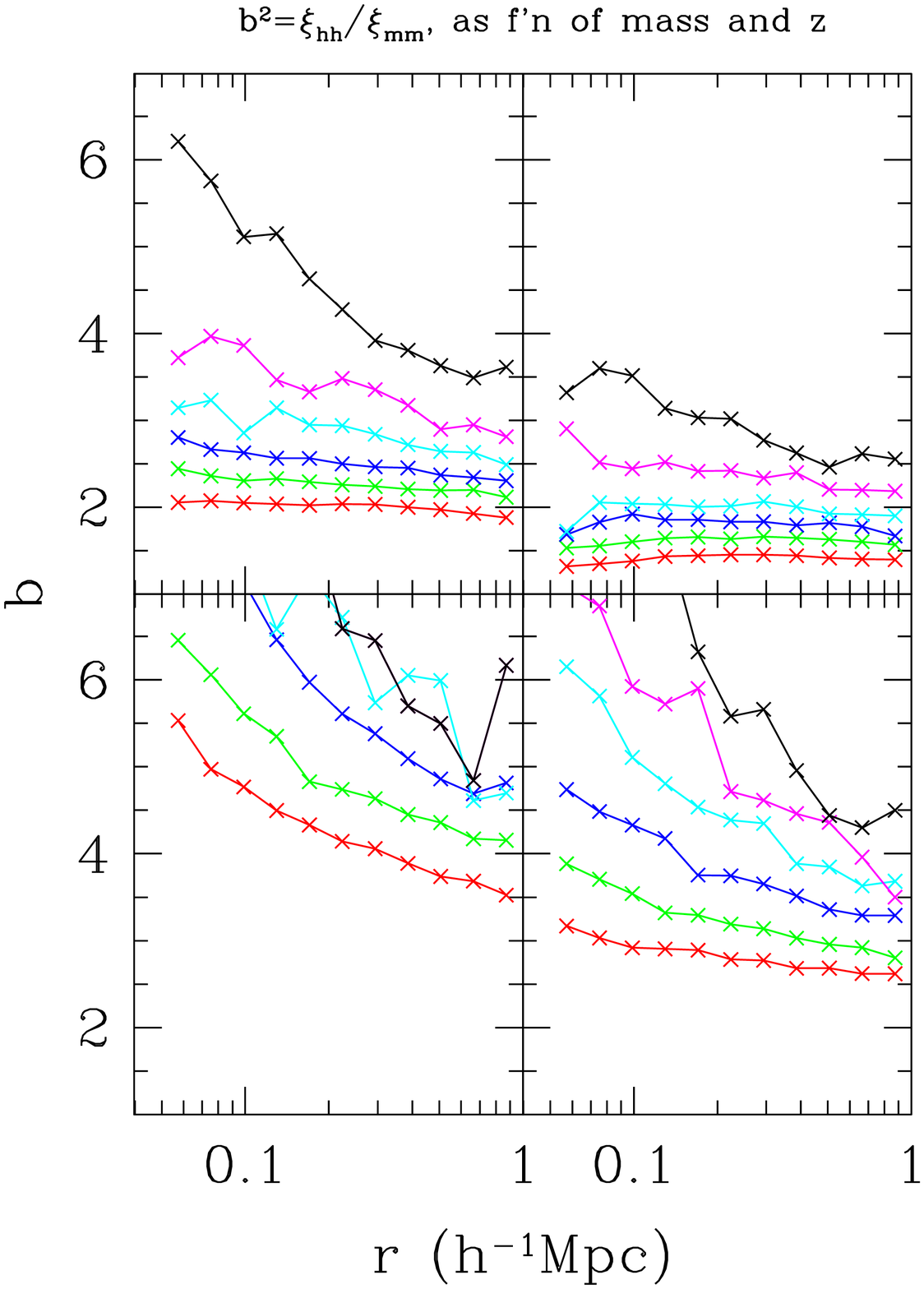}
\caption{shows the ratio of 
dark matter halo correlation function over the dark matter 
mass correlation function at four redshifts,
$z=6.0$ (top right panel),
$z=7.4$ (top left),
$z=9.1$ (bottom right)
and $z=11$ (bottom left).
In each panel six curves are shown 
for halos more massive than 
$10^{6}\msun$
$10^{6.5}\msun$
$10^{7}\msun$
$10^{7.5}\msun$
$10^{8}\msun$
and $10^{8.5}\msun$, respectively.
\label{f5}}
\end{figure}

\epsscale{1.0}
\begin{figure}
\plotone{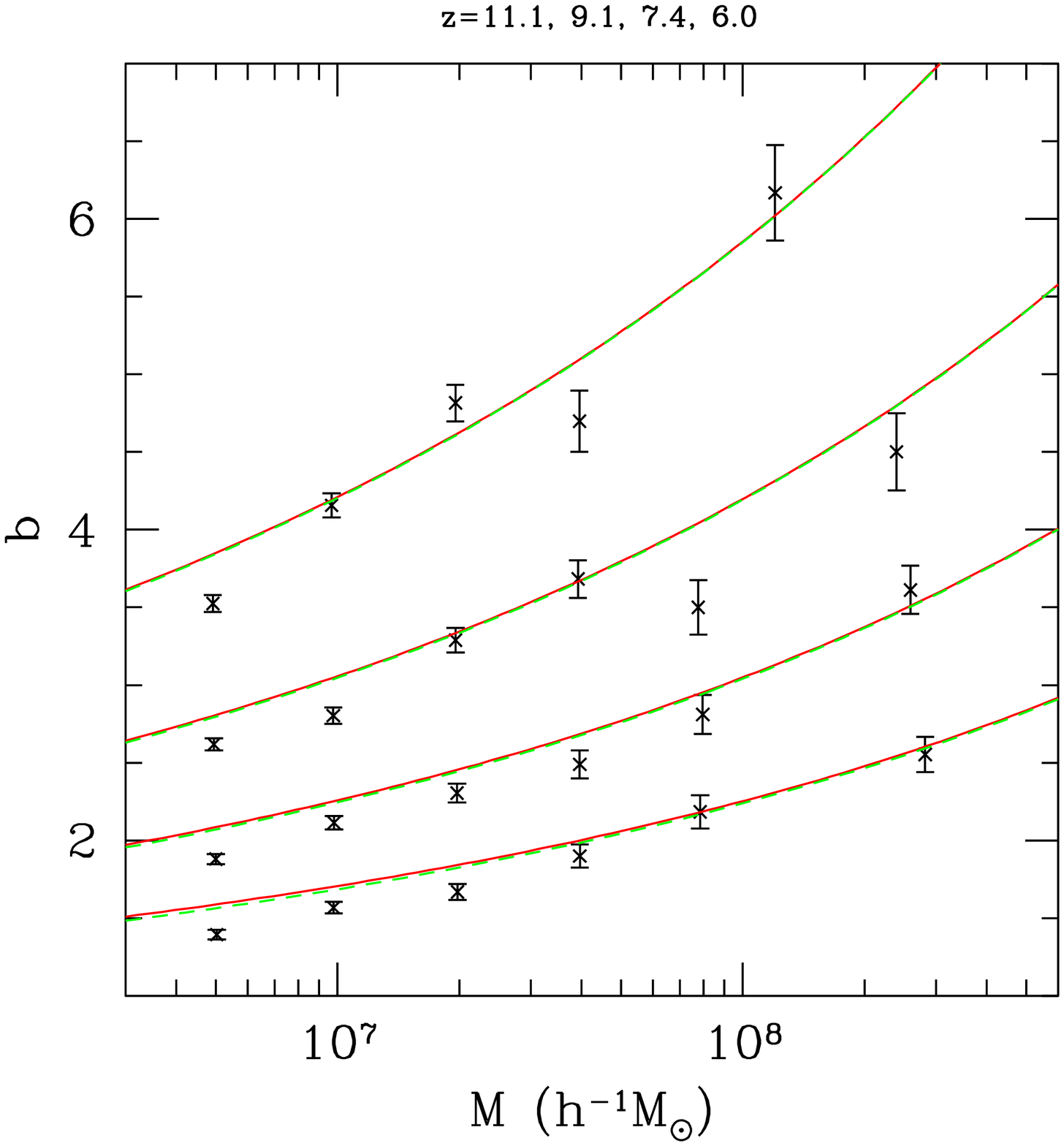}
\caption{
shows the bias as a function of halo mass at redshifts
$z=(6,7.4,9.1,11.1)$ at separation of $1h^{-1}$Mpc
as symbols ($x$).
The curves are computed using the analytic method of
Mo \& White (1996).
\label{f5}}
\end{figure}

We characterize the relative distribution of
halos over the total dark matter distribution
by the following relation:

\begin{equation}
{n_{h}\over <n_h>} = b(M,z)({\rho_{m}\over <\rho_m>})^{c(M,z)},
\end{equation}
\noindent
where $n_{h}$ and $<n_h>$
are the halo density and mean halo density;
$\rho_{m}$ and $<\rho_m>$
are the mass density and mean mass density;
$c(M,z)$ is fixed to be unity at ${\rho_{m}\over <\rho_m>}>1$.
This empirical fitting formula is motivated by
the found result that there appears to be a break 
in ${n_{h}\over <n_h>}$ at ${\rho_{m}\over <\rho_m>} \sim 1$.
Tables 2 and 3 list the parameters $b(M,z)$ and $c(M,z)$.
The smoothing length used here is 0.3 $h^{-1}$ Mpc. 
Figure 4 shows four typical cases to indicate the goodness
of the fitting formula. 
At ${\rho_{m}\over <\rho_m>}<1$ our results (Table 2) 
indicate ${n_{h}\over <n_h>} \propto ({\rho_{m}\over <\rho_m>})^{3-7}$,
a rather rapid drop.
As expected, the drop-off is more dramatic for larger halos,
as visible in Figure 2.
This implies that at $z>6$ halos are unlikely to be
found in underdense regions (on a scale of $\sim 0.3Mpc/h$).
The increase of $c(M,z)$ with redshift implies
that voids are emptier at higher redshifts.

Another way to characterize the relative distribution of 
dark matter halos over mass is to compute the 
ratio of the correlation functions, which are shown in Figure 5.
The correlation function $\xi(r)$ is calculated by counting
the number of pairs of either particles or halos at separation $r$
(using logarithmically spaced bins) and 
comparing that number to a Poisson distribution.
It can be seen that the bias falls in the range $2-6$,
with the trend that the more massive halos are more biased
and at a fixed mass halos at higher redshifts
are more biased, as expected.
Figure 6 recollects the information in Figure 5
and shows the bias as a function of halo mass at four different redshifts
at the scale chosen to be $1h^{-1}$Mpc.
The agreement between our computed results and that
using analytic method of Mo \& White (1996) is good, indicating that the latter
is valid for objects at scales and redshifts of concern here.

\subsection{Dark Matter Halo Density Profile}

\begin{figure*}
\includegraphics[width=2.8in]{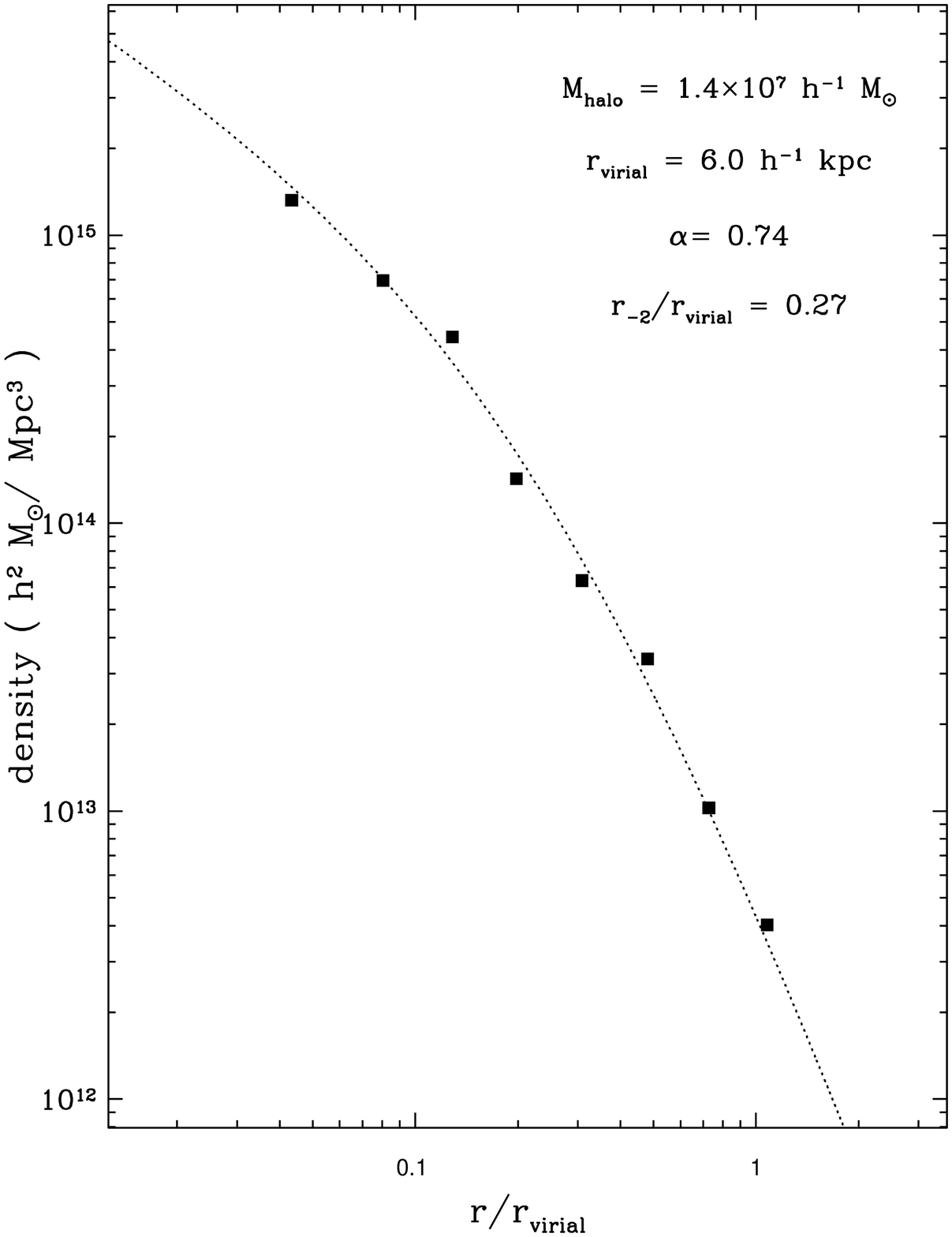}
\hspace{0.01in}
\includegraphics[width=2.8in]{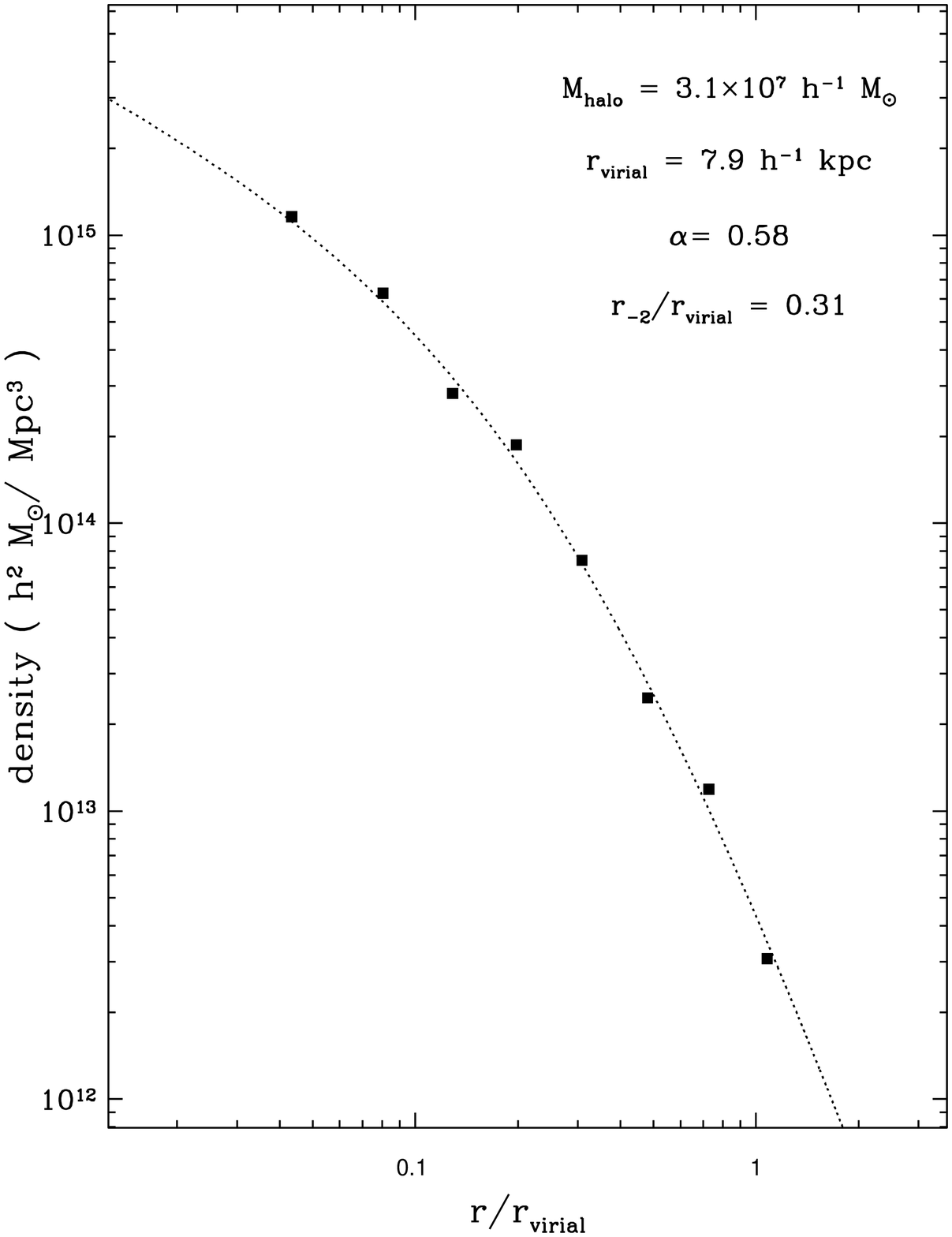}
\vskip 0.01in
\includegraphics[width=2.8in]{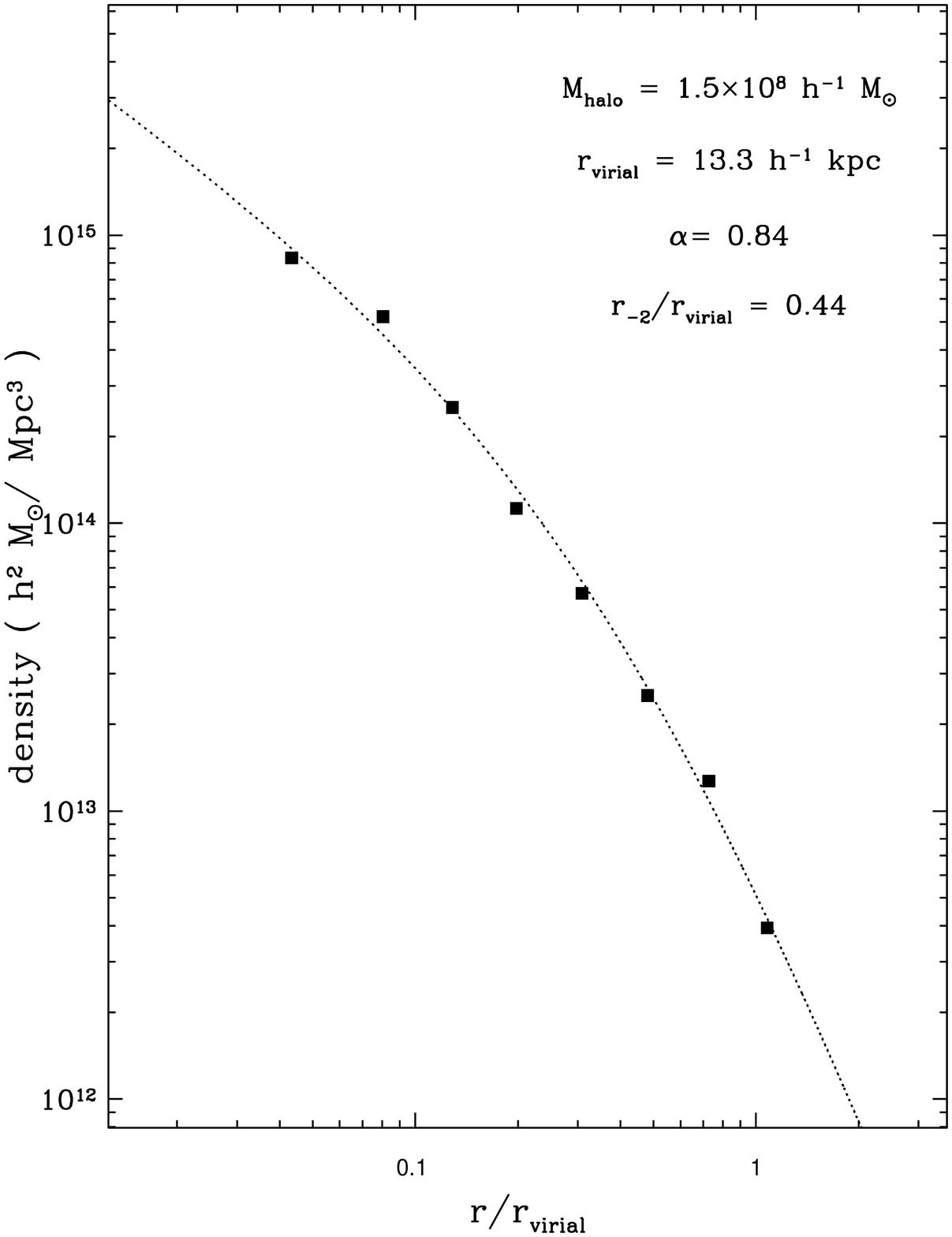}
\hspace{0.01in}
\includegraphics[width=2.8in]{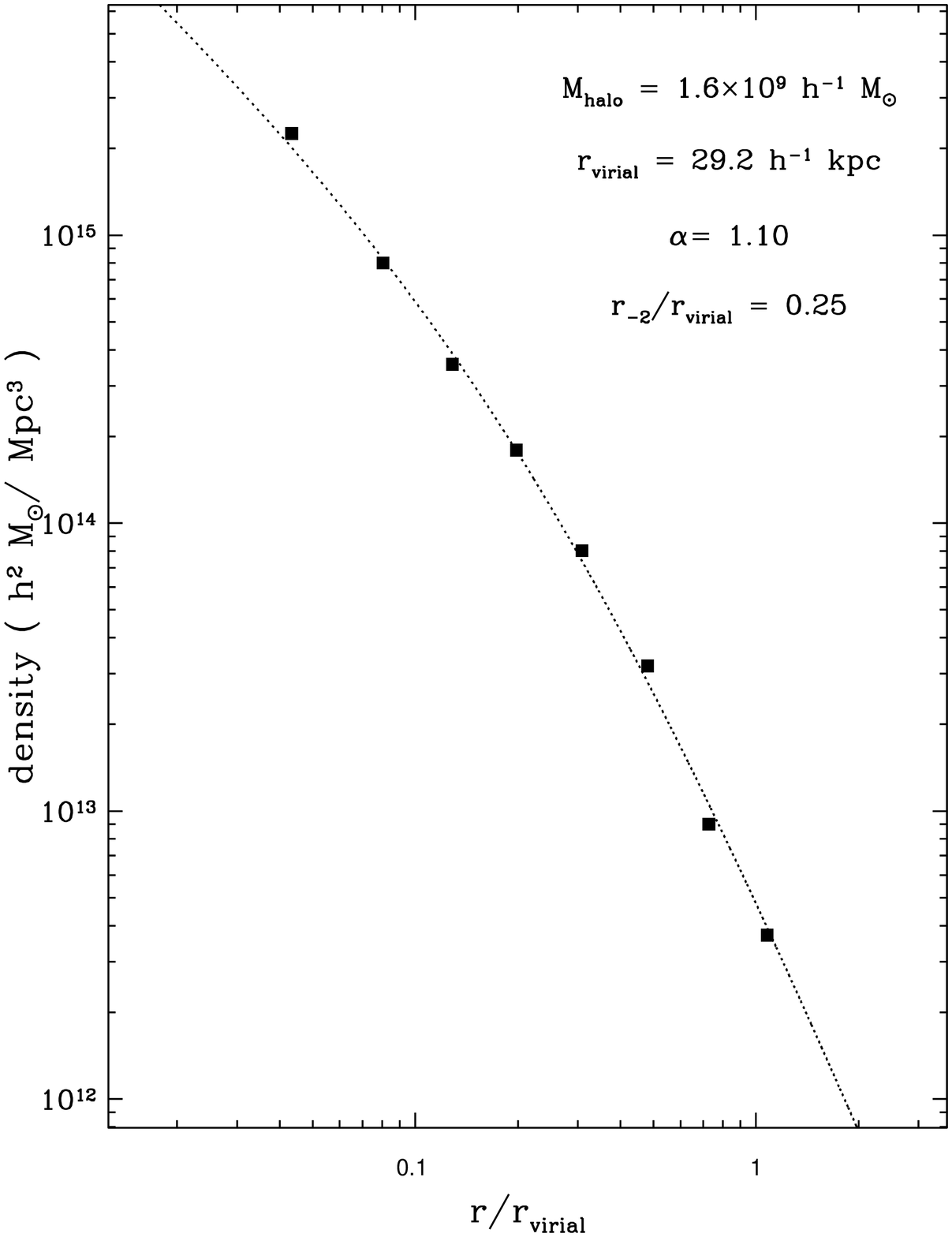}
\caption{Density profiles of four randomly selected halos over a mass range at 
$z=6$. The dotted lines represent the best-fit modified NFW relation given in 
Section 3. 
\label{f6}}
\end{figure*}

\begin{figure*}
\includegraphics[width=2.8in]{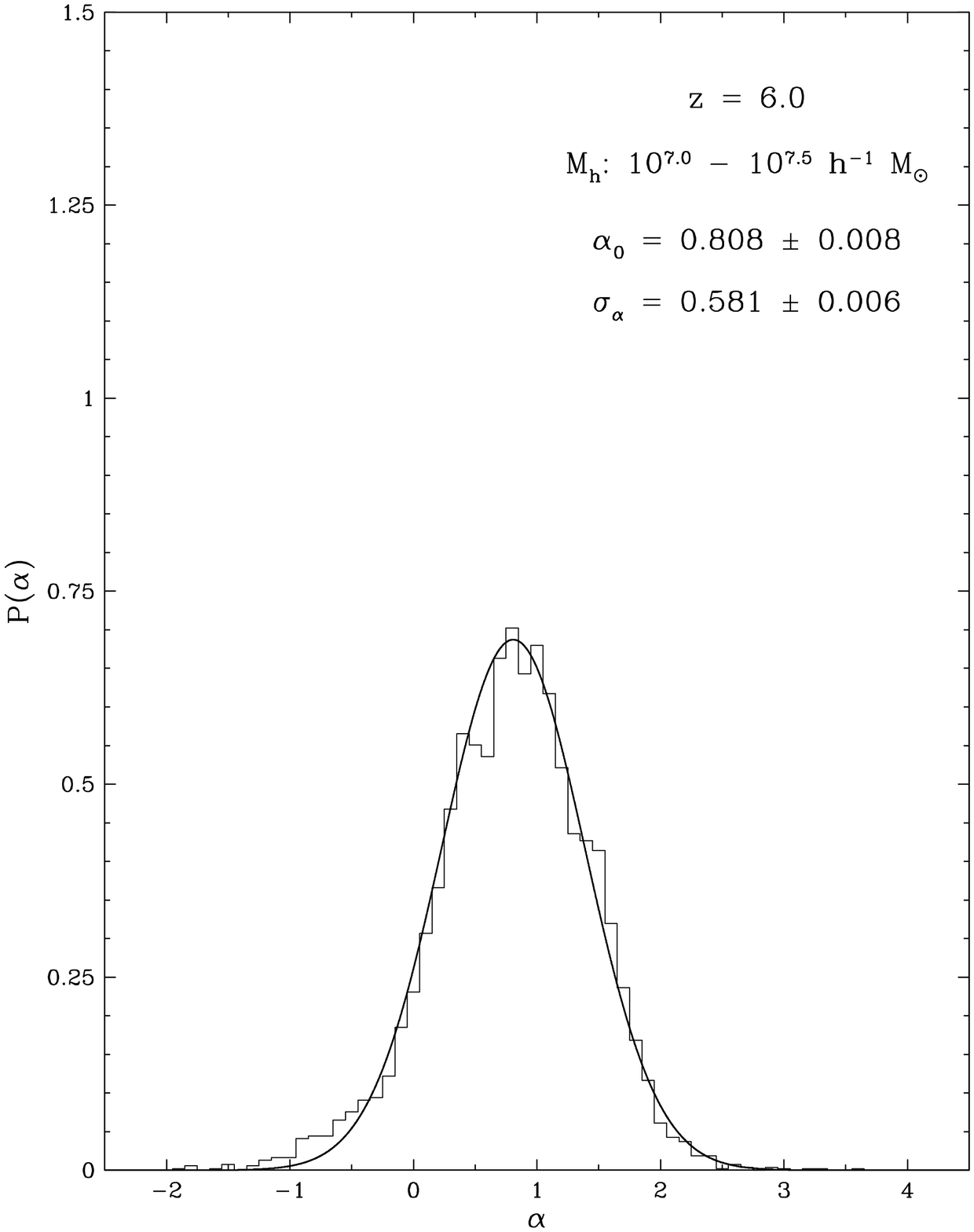}
\hspace{0.01in}
\includegraphics[width=2.8in]{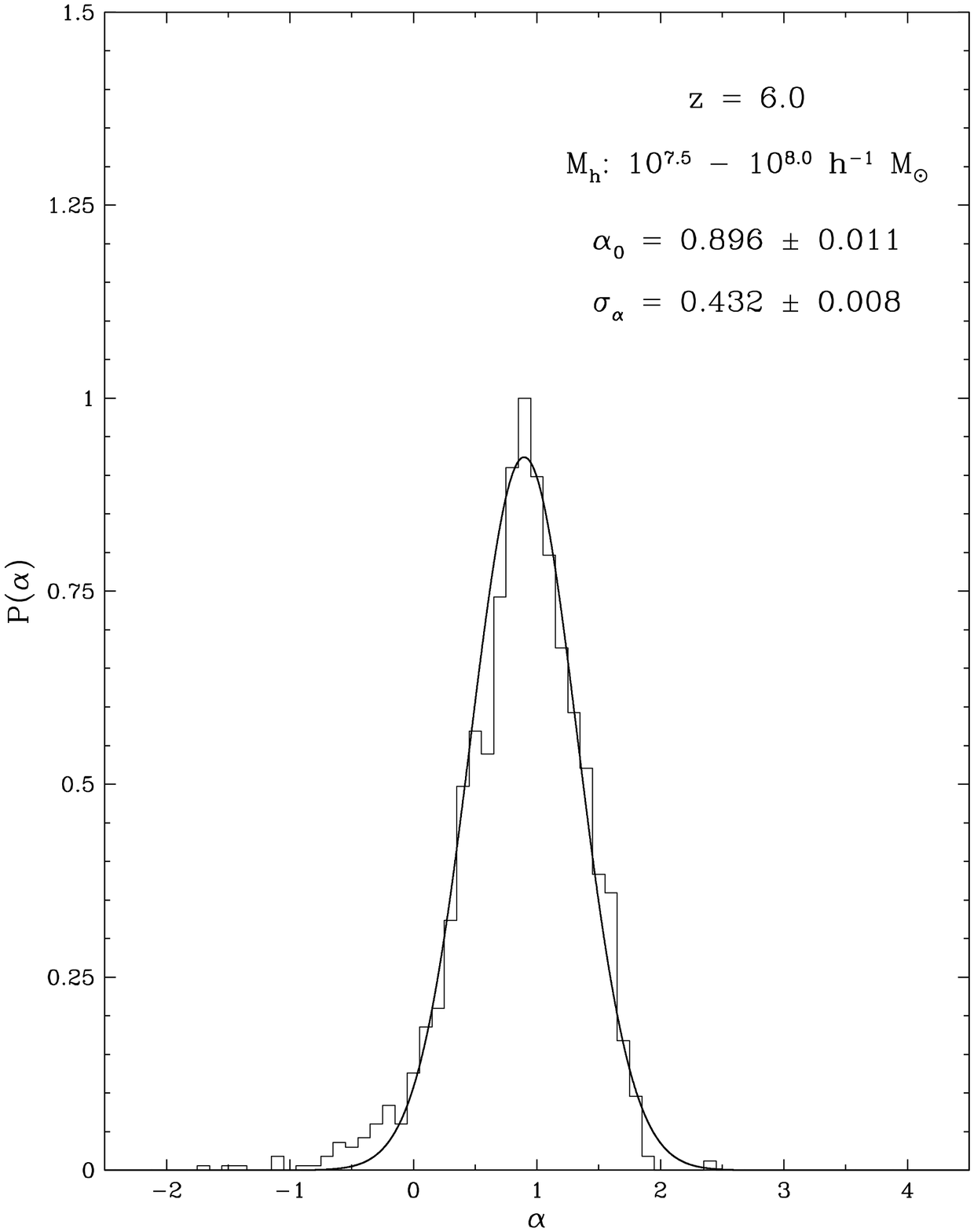}
\vskip 0.01in
\includegraphics[width=2.8in]{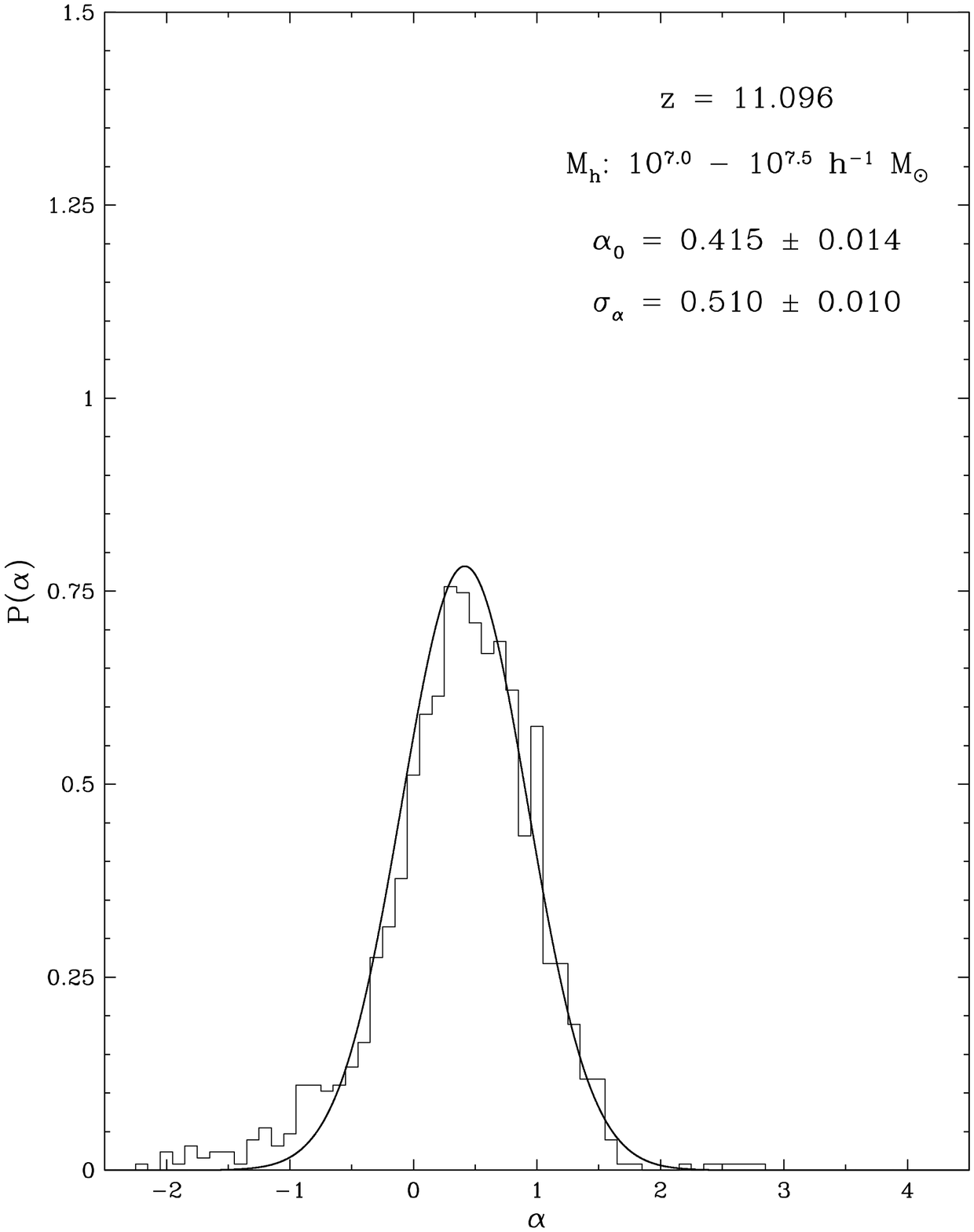}
\hspace{0.01in}
\includegraphics[width=2.8in]{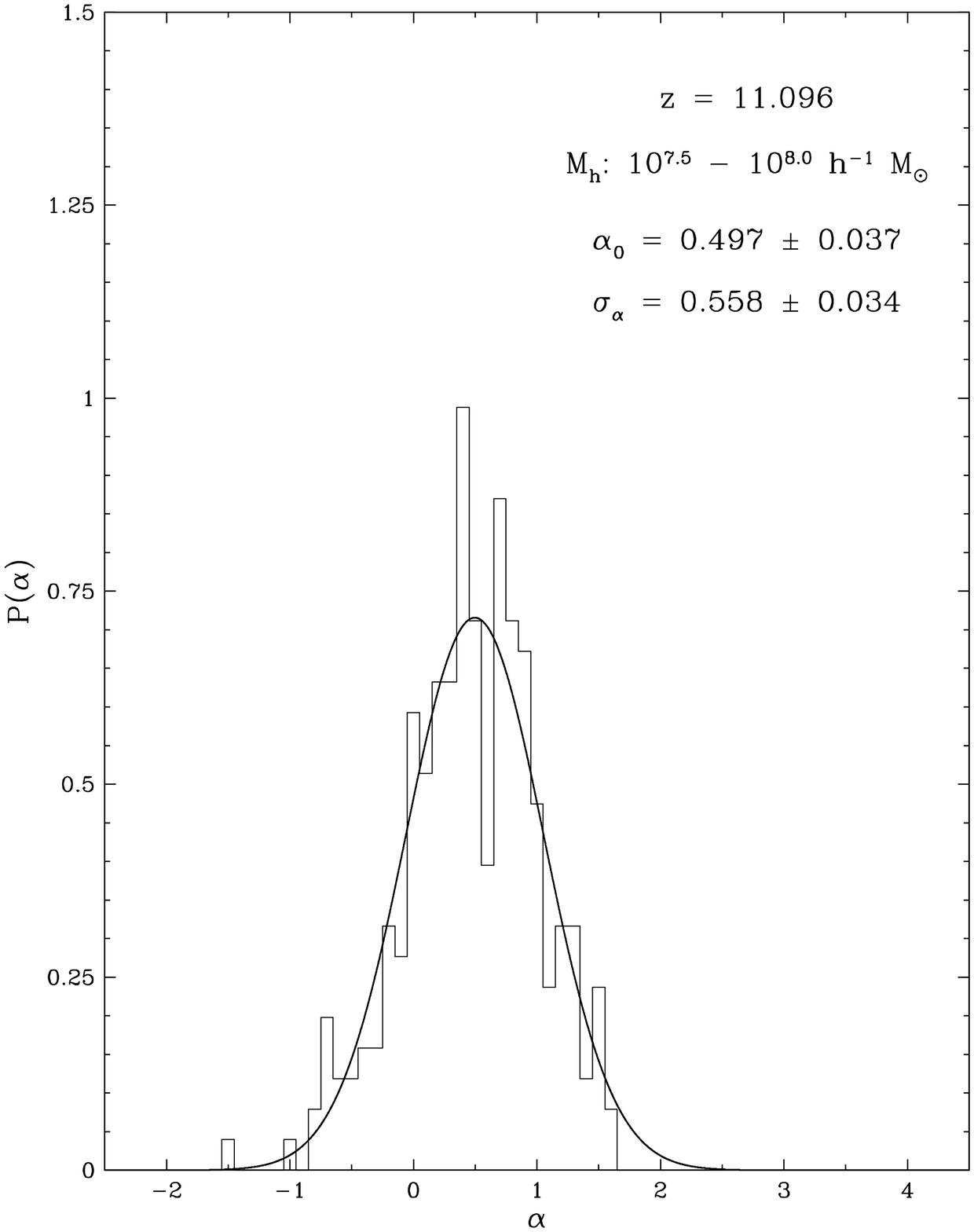}
\caption{The distributions of inner slope parameter $\alpha$ in the halo density profile fitting 
for four randomly selected cases upon different mass and redshift. The Gaussian fits 
are shown as smooth curves.
\label{f7}}
\end{figure*}
We use a variant fitting formula based on
the NFW (Navarro, Frenk, \& White 1997) profile:

\begin{equation}
\rho_{r} = {\rho_s\over ({r\over r_{-2}})^{\alpha} (1+{r\over r_{-2}})^{4-2\alpha}}.
\end{equation}
\noindent
Note that for NFW profile, $\alpha=1$.
An important difference, however, is the scaling radius used.
We use the radius where
the logarithmic slope of the density profile is $-2$, $r_{-2}$,
instead of the more conventional ``core" radius.
This is a two parameter fitting formula, 
$\alpha$ and $r_{-2}$,
while $\rho_s$ is a function of 
$\alpha$ and $r_{-2}$ at a fixed redshift, since
the overdensity interior to the virial radius $r_v$ is assumed to be known.
This fitting formula is intended for the range in radius $r\le r_v$ only.

We fit the density profile of each halo using the 
least squares method. Four randomly selected examples of such profiles 
along with the fitted curve using Equation (3) are shown in Figure 7,
indicating reasonable fits in all cases.
Both fitting parameters, $\alpha$ and $r_{-2}$,
however, display broad distributions.
We found that there is only weak correlation between
$\alpha$ and $r_{-2}$.
Figure 8 shows histograms for the distributions of $\alpha$,
for four typical cases.
We fit $\alpha$ distributions using a Gaussian distribution function:
\begin{equation}
P(\alpha) = {1\over \sqrt{2\pi} \sigma_\alpha}\exp{(-{(\alpha-\alpha_0)^2\over 2{\sigma_\alpha}^2})}.
\end{equation}
\noindent
The Gaussian fits are shown as smooth curves in Figure 8,
demonstrating that the proposed Gaussian fits are good.
Tables 4,5 list fitting parameters 
$\alpha_0$ and 
$\sigma_{\alpha_0}$, respectively. 
We recollect the data in Table 4 and  
shows in Figure 9 the median inner density slope  
as a function of dark matter halo different mass
at four different redshifts (symbols).
The curves in Figure 9 are empirical fits using the following formula
\begin{equation}
\alpha_0=0.75((1+z)/7.0)^{-1.25}(M/10^7\msun)^{0.11(1+z)/7.0}.
\end{equation}
\noindent
It is seen that this fitting formula provides a reasonable fit
for the simulated halos.

\epsscale{0.8}
\begin{figure}
\plotone{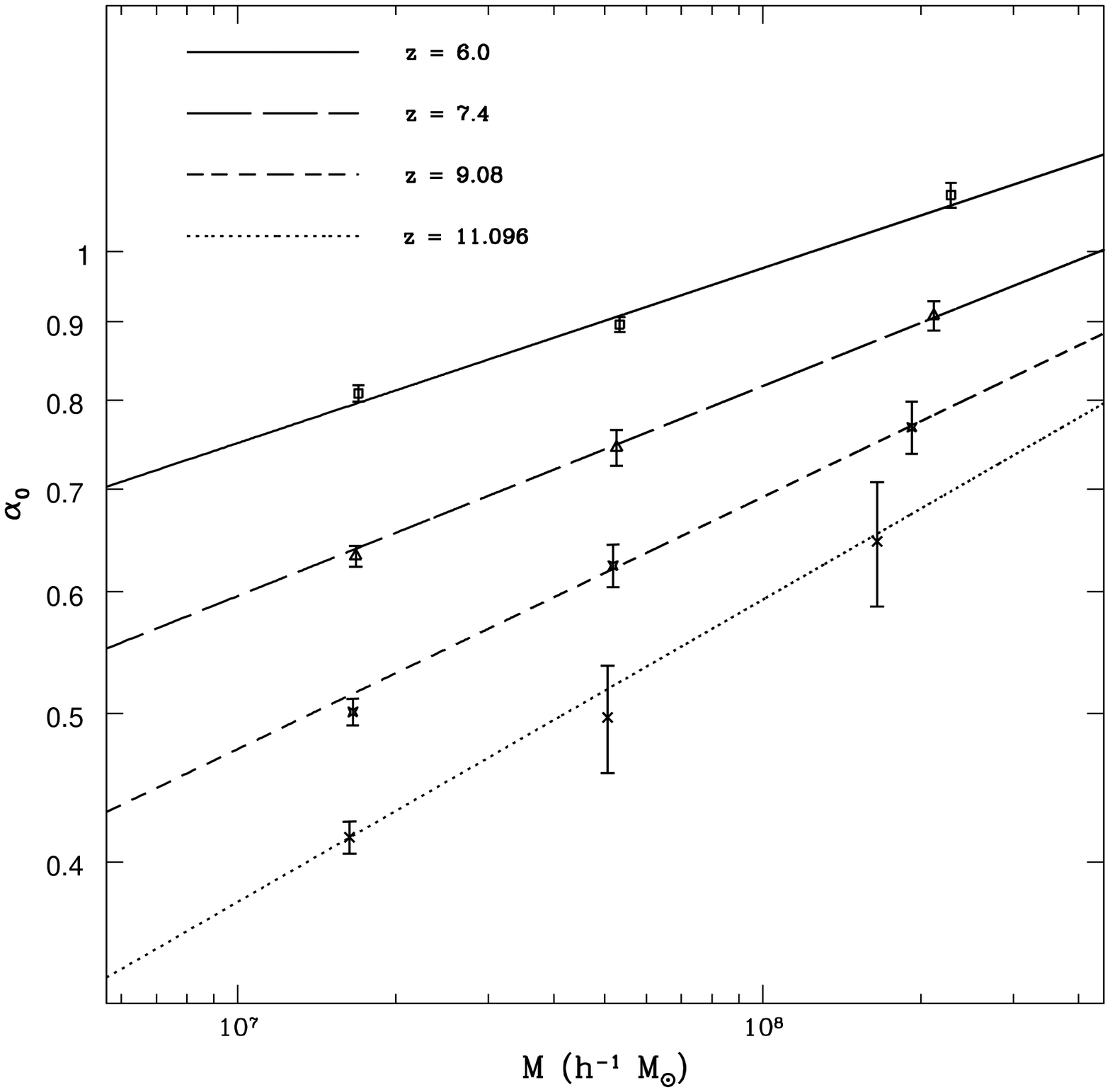}
\caption{shows the median inner density slope  
as a function of dark matter halo different mass
at four different redshifts (symbols).
The curves are fits (Equation 5).
\label{f9}}
\end{figure}

\begin{deluxetable}{rrrrrrrr} 
\tablecolumns{5} 
\tablewidth{0pc} 
\tablecaption{Density Profile : $\alpha_0(M,z)$} 
\tablehead{ 
\colhead{Halo Mass ($h^{-1}$ M$_\odot$) } & \colhead{\ \ \ z=6.0 \ \ }   & \colhead{\ \ z=7.4 \ }  & \colhead{z=9.08 \ } & \colhead{\ \ z=11.096}}
\startdata 
10$^{7.0}$ - 10$^{7.5}$  \ \ \ \ \ & \ 0.81$\pm$0.01 & \ \ \ 0.63$\pm$0.01 \ \ & \ 0.50$\pm$0.01 \ & 
\ \ 0.42$\pm$0.01 \\ 
10$^{7.5}$ - 10$^{8.0}$ \ \ \ \ \ & \ 0.90$\pm$0.01 & \ \ \ 0.75$\pm$0.02 \ \ & \ 0.62$\pm$0.02 \ & 
\ \ 0.50$\pm$0.04 \\ 
$>$ 10$^{8.0}$ \ \ \ \ \ \ \ \ \ & \ 1.09$\pm$0.02 & \ \ \ 0.91$\pm$0.02 \ \ & \ 0.77$\pm$0.03 \ & 
\ \ 0.65$\pm$0.06 \\ 
\hline
\enddata 
\end{deluxetable} 

\begin{deluxetable}{rrrrrrrr} 
\tablecolumns{5} 
\tablewidth{0pc} 
\tablecaption{Density Profile : $\sigma_{\alpha}(M,z)$} 
\tablehead{ 
\colhead{Halo Mass ($h^{-1}$ M$_\odot$)} & \colhead{\ \ \ z=6.0 \ \ }   & \colhead{\ \ z=7.4 \ }  & \colhead{z=9.08 \ } & \colhead{\ \ z=11.096}}
\startdata 
10$^{7.0}$ - 10$^{7.5}$ \ \ \ \ \ & \ 0.58$\pm$0.01 & \ \ \ 0.54$\pm$0.01 \ \ & \ 0.52$\pm$0.01 \ & 
\ \ 0.51$\pm$0.01 \\ 
10$^{7.5}$ - 10$^{8.0}$ \ \ \ \ \ & \ 0.43$\pm$0.01 & \ \ \ 0.45$\pm$0.01 \ \ & \ 0.46$\pm$0.02 \ & 
\ \ 0.56$\pm$0.03 \\ 
$>$ 10$^{8.0}$ \ \ \ \ \ \ \ \ \ & \ 0.41$\pm$0.01 & \ \ \ 0.41$\pm$0.02 \ \ & \ 0.41$\pm$0.03 \ & 
\ \ 0.38$\pm$0.06 \\ 
\hline
\enddata 
\end{deluxetable} 

\begin{deluxetable}{rrrrrrrr} 
\tablecolumns{5} 
\tablewidth{0pc} 
\tablecaption{Density Profile : $r_{-2}^0$(M,z)} 
\tablehead{ 
\colhead{Halo Mass ($h^{-1}$ M$_\odot$)} & \colhead{\ \ \ z=6.0 \ \ }   & \colhead{\ \ z=7.4 \ }  & \colhead{z=9.08 \ } & \colhead{\ \ z=11.096}}
\startdata 
10$^{7.0}$ - 10$^{7.5}$ \ \ \ \ \ & \ 0.35$\pm$0.01 & \ \ \ 0.40$\pm$0.01 \ \ & \ 0.43$\pm$0.01 \ & 
\ \ 0.44$\pm$0.01 \\ 
10$^{7.5}$ - 10$^{8.0}$ \ \ \ \ \ & \ 0.35$\pm$0.01 & \ \ \ 0.38$\pm$0.01 \ \ & \ 0.39$\pm$0.01 \ & 
\ \ 0.37$\pm$0.01 \\ 
$>$ 10$^{8.0}$  \ \ \ \ \ \ \ \ \ & \ 0.34$\pm$0.01 & \ \ \ 0.35$\pm$0.01 \ \ & \ 0.34$\pm$0.01 \ & 
\ \ 0.34$\pm$0.02 \\ 
\hline
\enddata 
\end{deluxetable} 

\begin{deluxetable}{rrrrrrrr} 
\tablecolumns{5} 
\tablewidth{0pc} 
\tablecaption{Density Profile : $\sigma_{r_{-2}}$(M,z)} 
\tablehead{ 
\colhead{Halo Mass ($h^{-1}$ M$_\odot$)} & \colhead{\ \ \ z=6.0 \ \ }   & \colhead{\ \ z=7.4 \ }  & \colhead{z=9.08 \ } & \colhead{\ \ z=11.096}}
\startdata 
10$^{7.0}$ - 10$^{7.5}$ \ \ \ \ \ & \ 0.34$\pm$0.01 & \ \ \ 0.32$\pm$0.01 \ \ & \ 0.29$\pm$0.01 \ & 
\ \ 0.27$\pm$0.01 \\ 
10$^{7.5}$ - 10$^{8.0}$ \ \ \ \ \ & \ 0.31$\pm$0.01 & \ \ \ 0.31$\pm$0.01 \ \ & \ 0.27$\pm$0.01 \ & 
\ \ 0.27$\pm$0.01 \\ 
$>$ 10$^{8.0}$  \ \ \ \ \ \ \ \ \ & \ 0.33$\pm$0.01 & \ \ \ 0.25$\pm$0.01 \ \ & \ 0.26$\pm$0.02 \ & 
\ \ 0.27$\pm$0.05 \\ 
\hline
\enddata 
\end{deluxetable}

Figure 10 shows histograms for the distributions of $r_{-2}$
for four typical cases.
We fit $r_{-2}$ distributions using a lognormal function:
\begin{equation}
P(r_{-2}) = {1\over r_{-2}\sqrt{2\pi} \sigma_{r_{-2}}}\exp{(-{(\ln r_{-2} -\ln r_{-2}^0)^2\over 2\sigma_{r_{-2}}})},
\end{equation}
\noindent
which are seen to provide reasonable fits to the data in Figure 10.
Tables 6,7 list fitting parameters 
$r_{-2}^0$ and 
$\sigma_{r_{-2}}$, respectively. 
We find no visible correlation between $\alpha$ and $r_{-2}$,
as shown in Figure 11.

\begin{figure*}
\includegraphics[width=2.8in]{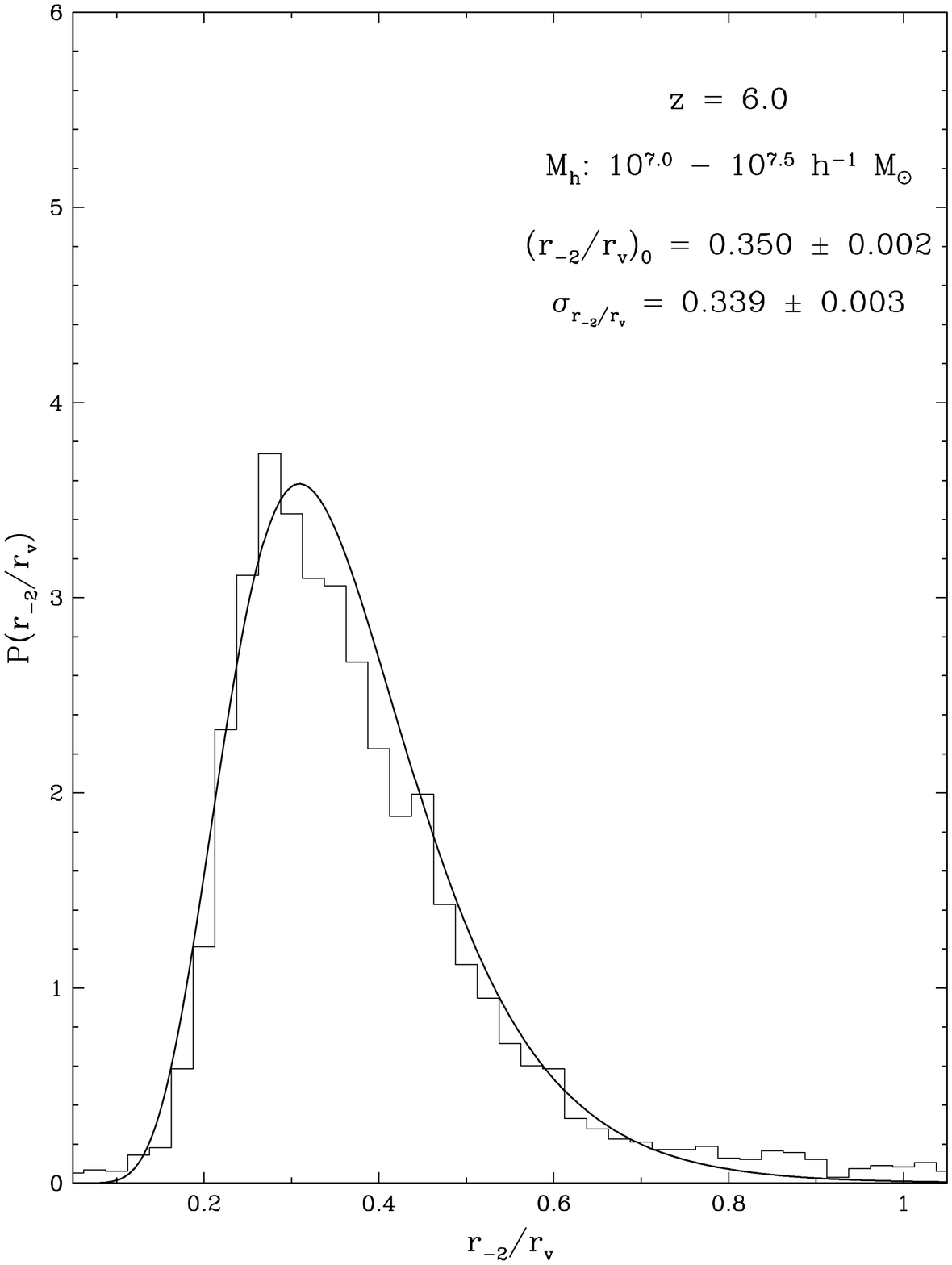}
\hspace{0.01in}
\includegraphics[width=2.8in]{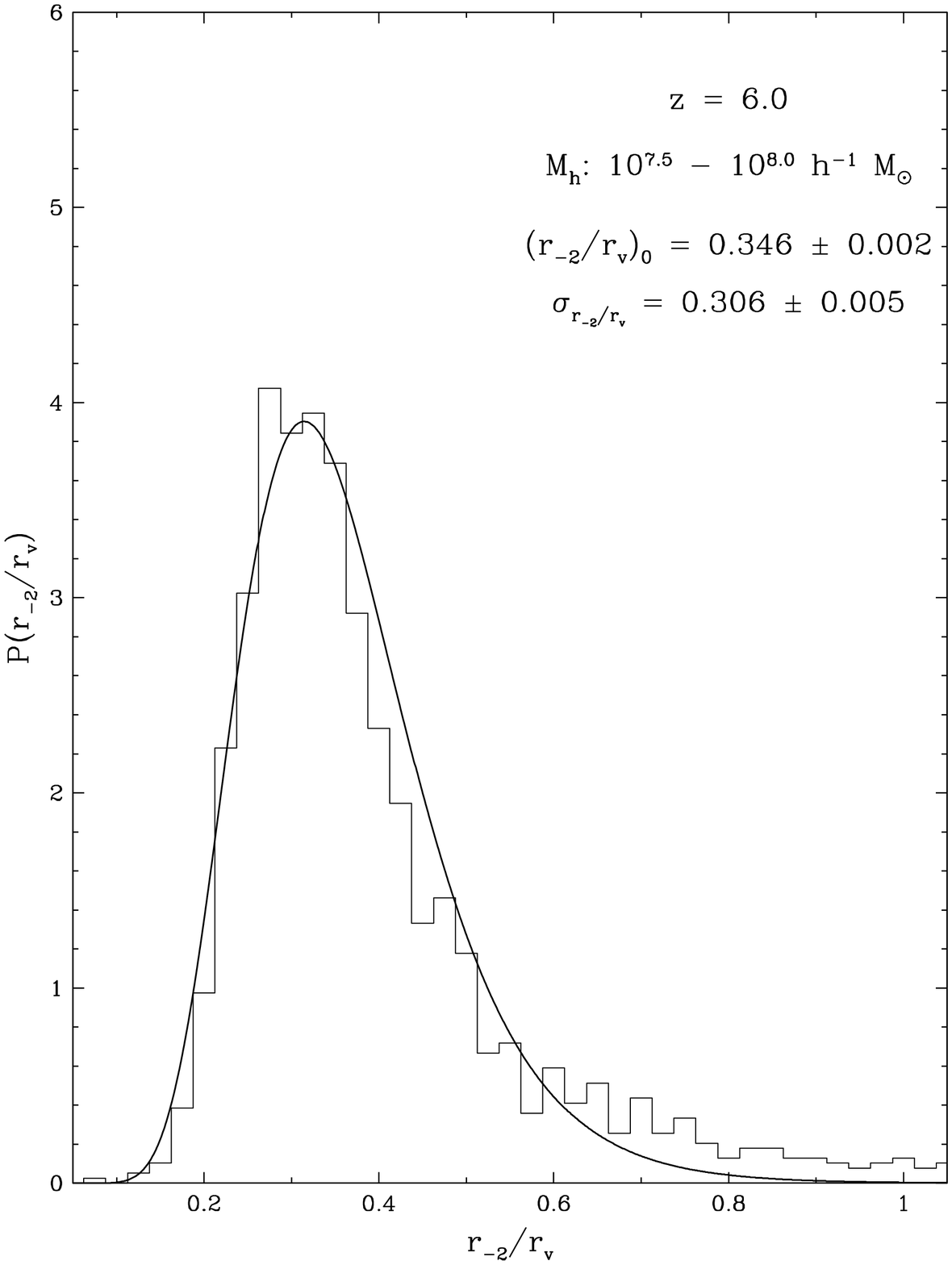}
\vskip 0.01in
\includegraphics[width=2.8in]{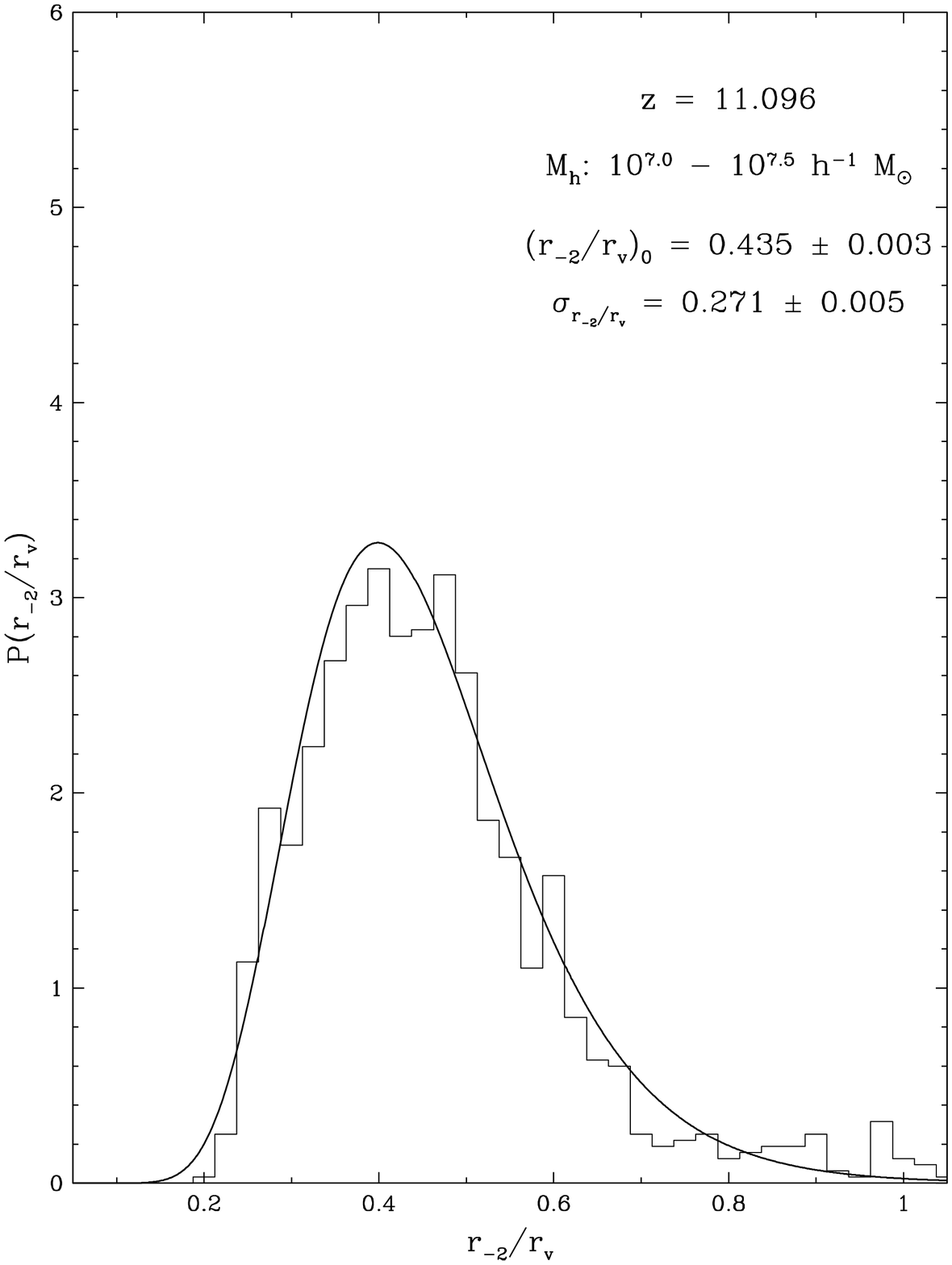}
\hspace{0.01in}
\includegraphics[width=2.8in]{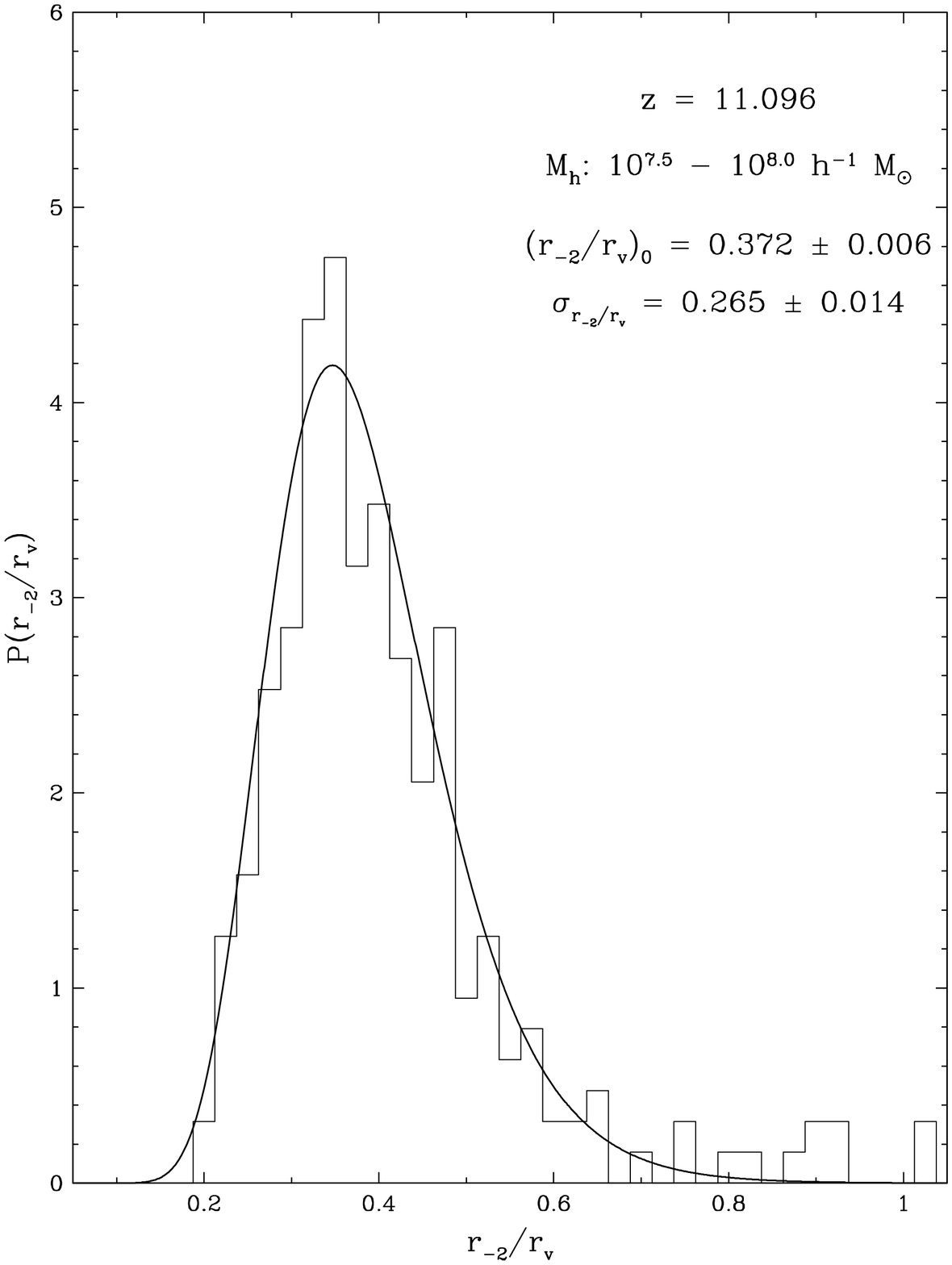}
\caption{The distributions of radius parameter $r_{-2}$ in the halo density profile fitting 
for four randomly selected cases upon different mass and redshift. The lognormal fits 
are shown as smooth curves.
\label{f8}}
\end{figure*}

An important point to note is that the average slope 
of the density profiles of small halos
ranges from $1.0$ to $0.4$ from $z=6$ to $z=11$
(Table 4 and Figure 8), which is somewhat shallower
than the universal density profile found by Navarro \etal (1997).
Our results at $z=9-11$ are in good agreement with Ricotti (2003),
who simulated a $1h^{-1}$Mpc box for  small halos at $z=10$.
If the theoretical argument for the dependence of 
halo density profile on the slope of the initial density fluctuation
power spectrum (Syer \& White  1988; Subramanian, Cen, \& Ostriker 2000)
is correct, as adopted by Ricotti (2003) to explain 
the dependence of inner density slope on halo mass,
it then follows that the neglect of density fluctuations 
on scales larger than his box size of $1h^{-1}$Mpc in Ricotti (2003)
would make the density profiles of the halos
in his simulation somewhat shallower than they should be in the CDM model.
Thus, the difference in the box size ($1h^{-1}$Mpc in Ricotti 2003
versus $4h^{-1}$Mpc for our simulation box)
would have expected to result in a slightly steeper inner slope
in our simulation, which is indeed the case.

Another point to note, which is not new but not widely known,
is that there is a large dispersion in the inner slope
of order $0.5$ due to the intrinsically stochastic nature of halo assembly.
This was found earlier by Subramanian \etal (2000).
Therefore, while a ``universal" profile is informative
in characterizing the mode,
a dispersion would be needed to give a full account.
This is particularly important for applications where
the dependence on the inner slope is very strong,
e.g., strong gravitational lensing.
More relevant for our case of small halos
is that, for example, the fraction of small halos
with inner slope close to zero (i.e., flat core)
is non-negligible at $z=6$.
A proper statistical comparison with observations
of local dwarf galaxies, however, is not possible with the current
simulation without evolving small galaxies to $z=0$.

\epsscale{1.0}
\begin{figure}
\includegraphics[width=1.0\hsize,angle=270]{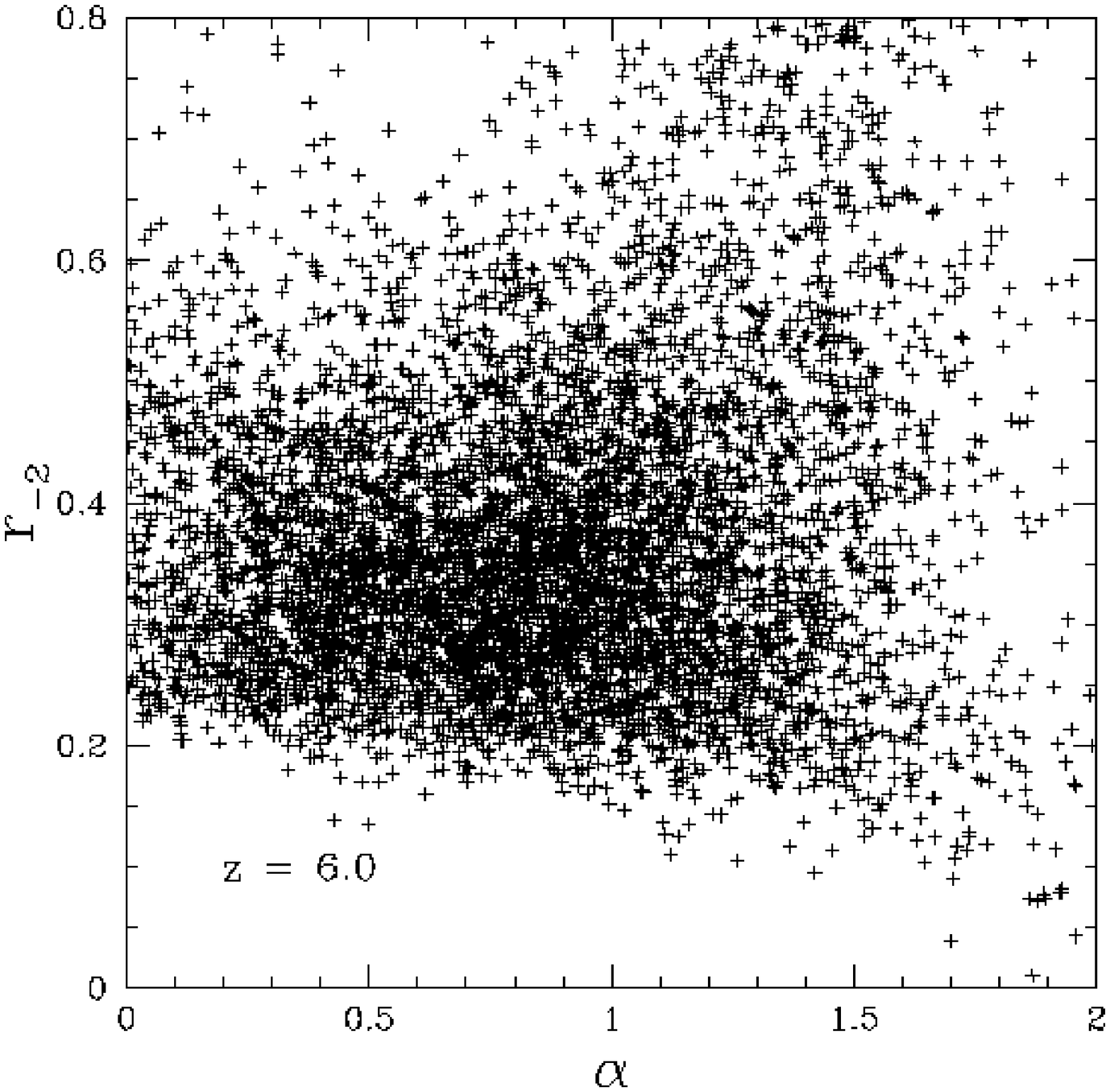}
\caption{
A scatter plot of the inner slope of the density profile,  $\alpha$,
versus $r_{-2}$ for all the halos with $M>10^{6.5} \ h^{-1} M_\odot$ at $z=6$. 
\label{f10}}
\end{figure}

\epsscale{1.0}
\begin{figure}
\includegraphics[width=1.0\hsize,angle=270]{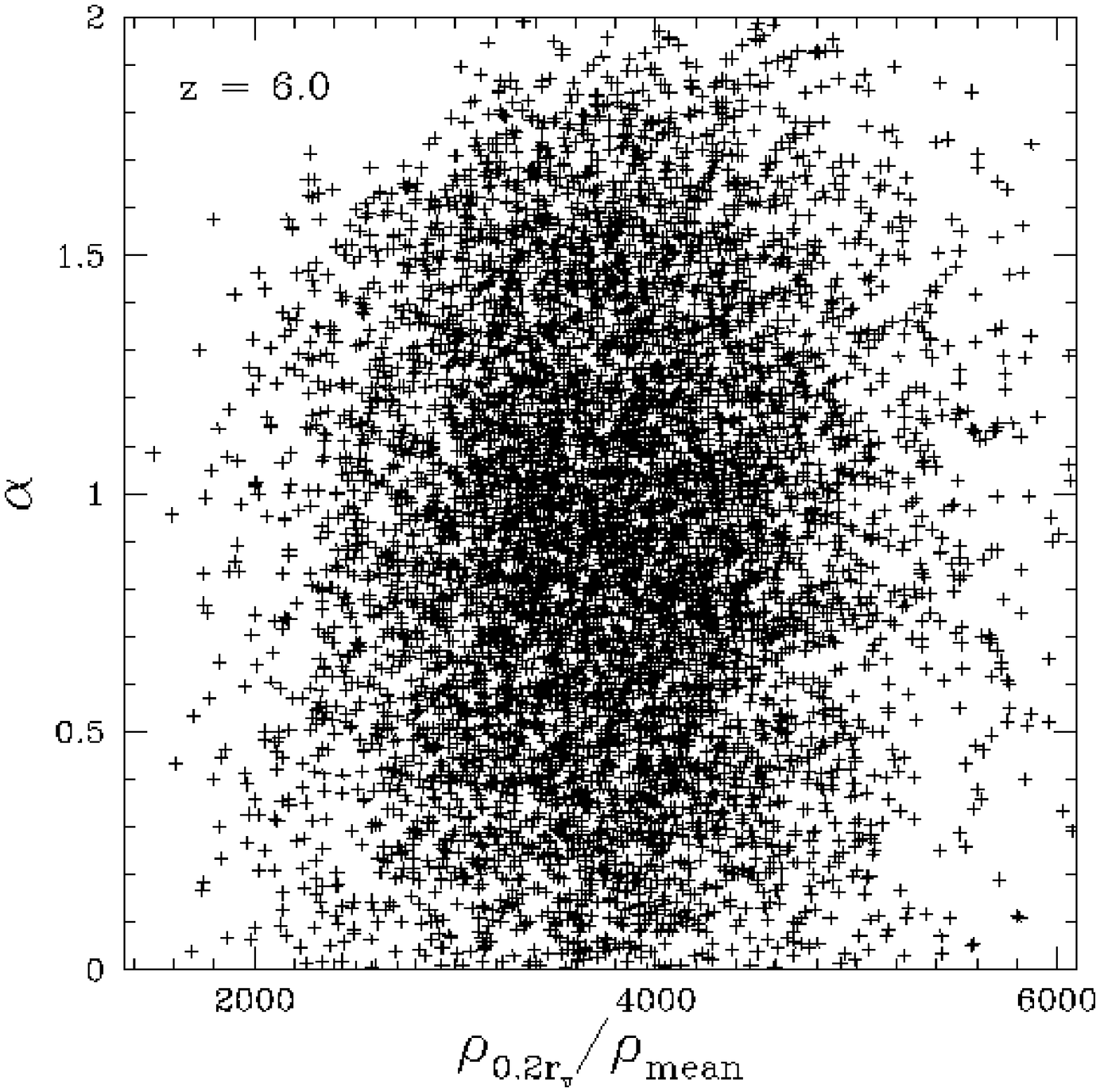}
\caption{
A scatter plot of the inner slope of the density profile,  $\alpha$,
versus the halo central density, defined as the density at $r<0.2r_{v}$. 
for all the halos with $M>10^6 \ h^{-1} M_\odot$ at $z=6$. 
\label{f10a}}
\end{figure}

\epsscale{1.0}
\begin{figure}
\includegraphics[width=1.0\hsize,angle=0]{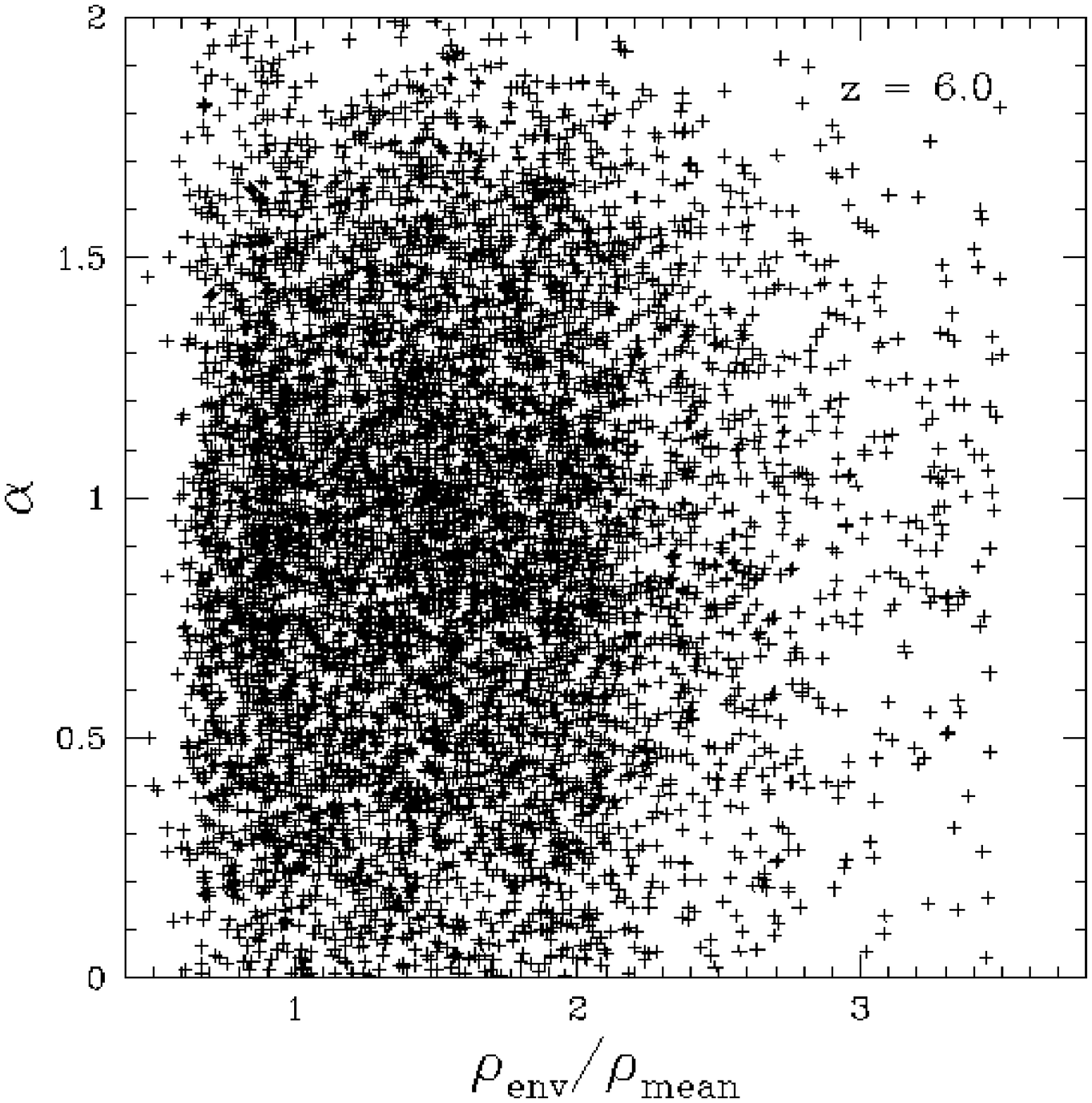}
\caption{
A scatter plot of the inner slope of the density profile,  $\alpha$,
versus the environmental density, defined as the dark matter density
smoothed by a gaussian window of radius $0.3h^{-1}$Mpc,
for all the halos with $M>10^6 \ h^{-1} M_\odot$ at $z=6$. 
\label{f10b}}
\end{figure}

Also worthwhile is to understand whether or not there is some dependence
of the inner slope  of the density profile on
the central density of a halo or the environmental density
where a halo sits.
In Figures 12 and 13 we show the correlation
between the inner slope of the density profile, $\alpha$,
and the central density of the halo,
and between $\alpha$ and the environmental density, respectively.
We find no visible correlations between either pair of quantities.
But a relationship between halo shape and environment might have been
missed due to stochastic variations.
Thus, we have checked to see if deviation
$\Delta\alpha$, between computed and predicted (based on mass and epoch
using equation 5) slope exists.
Figure 14 shows this and no correlation is seen.
We note that the abundances of halos in the mass range of interest here
are on the rise in the redshift range considered (Lacey \& Cole 1993),
as evident in Figure 3.
During this period halos considered the inner density profiles
of halos steepen with time, consistent with the 
increase of logarithmic slope of the power spectrum with time,
that corresponds to the evolving nonlinear mass scale.
However, at some lower redshift not probed here, 
low mass halos will cease to form.
Subsequently, the evolution of halo density profile 
may show distinct features and some conceivable correlations, not seen
in Figures 12 and 13, may show up.
We will study this issue separately.

\epsscale{1.0}
\begin{figure}
\plotone{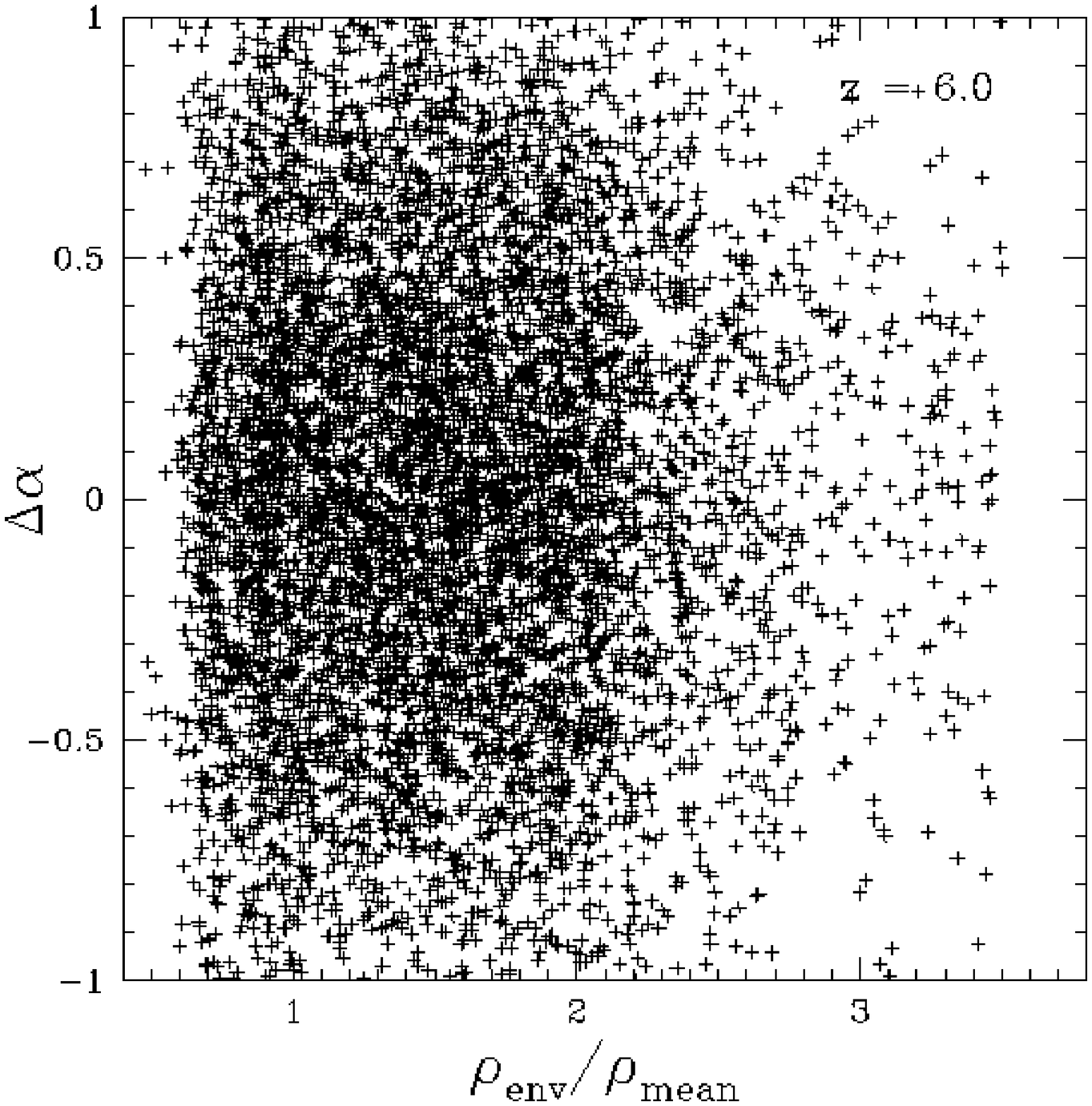}
\caption{
A scatter plot of 
the difference between measured inner slope 
of the density profile,  $\alpha$,
and the fitted inner slope using equation (5),
versus the environmental density.
\label{f10b}}
\end{figure}

Since the central density of a halo may be considered
a good proxy for the formation redshift of the central region,
the non-correlation between $\alpha$ and the central density
indicates that, for halos of question here,
the subsequent process of accretion of mass onto halos
is largely a random process,
independent of the density of the initial central ``seed".
The fact that halos of different masses show a comparable range 
of central density (not shown in figures)
suggest that halos of varying masses form nearly simultaneously,
dictated by the nature of the cold dark matter power spectrum
at the high-k end; i.e., density fluctuations on those scales 
involved here depend weakly (logarithmically) on the mass.
The non-correlation between the inner slope of the halo density profile
and the environmental density may be interpreted in the following way.
One may regard regions of different overdensities as 
local mini-universes of varying density parameters.
The independence of the inner slope on local density
is thus consistent with published results that
the halo density profiles in universes of different $\Omega_M$
do not significantly vary.

\subsection{Spin Parameter For Dark Matter Halos}

\begin{figure*}
\includegraphics[width=2.8in]{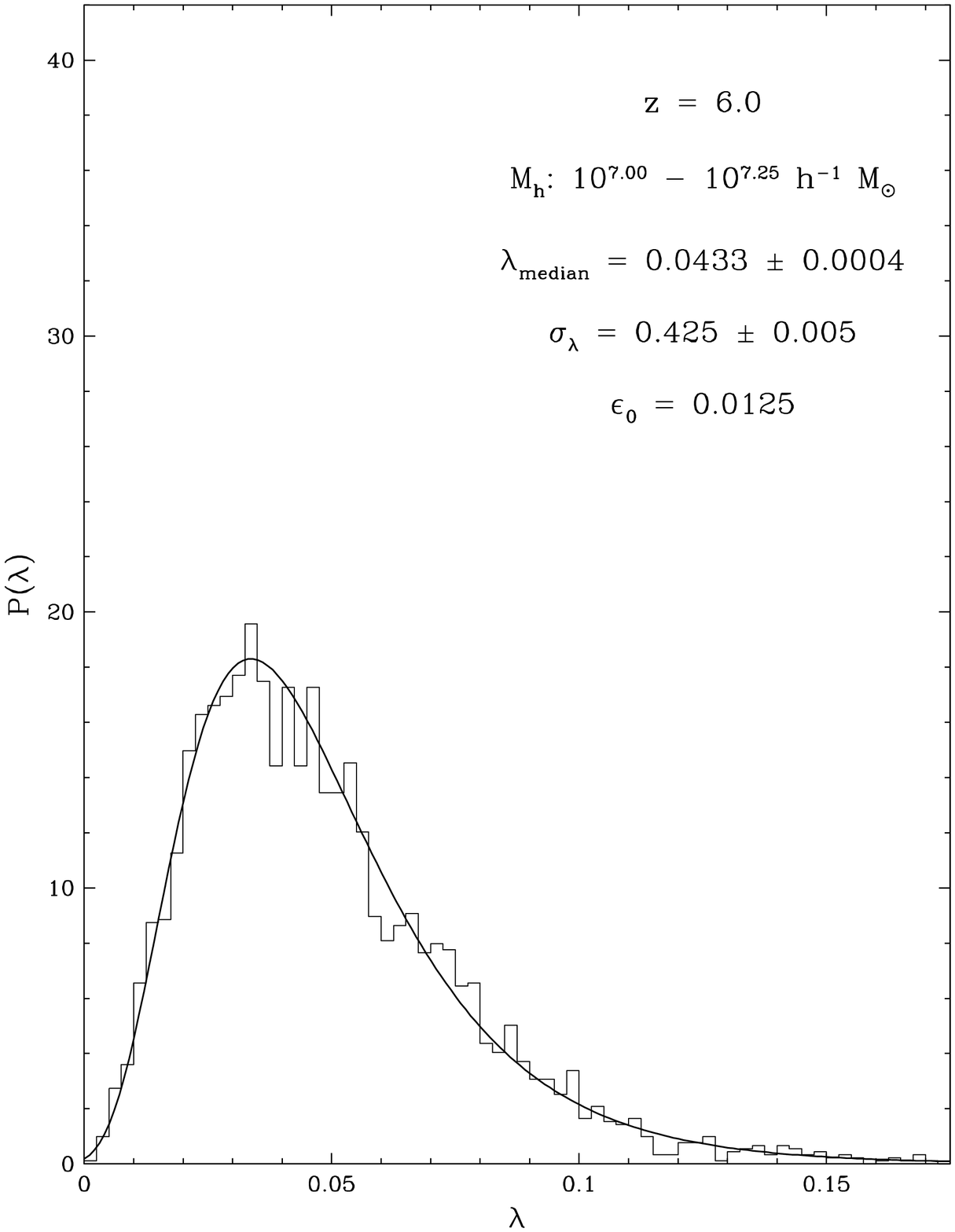}
\hspace{0.01in}
\includegraphics[width=2.8in]{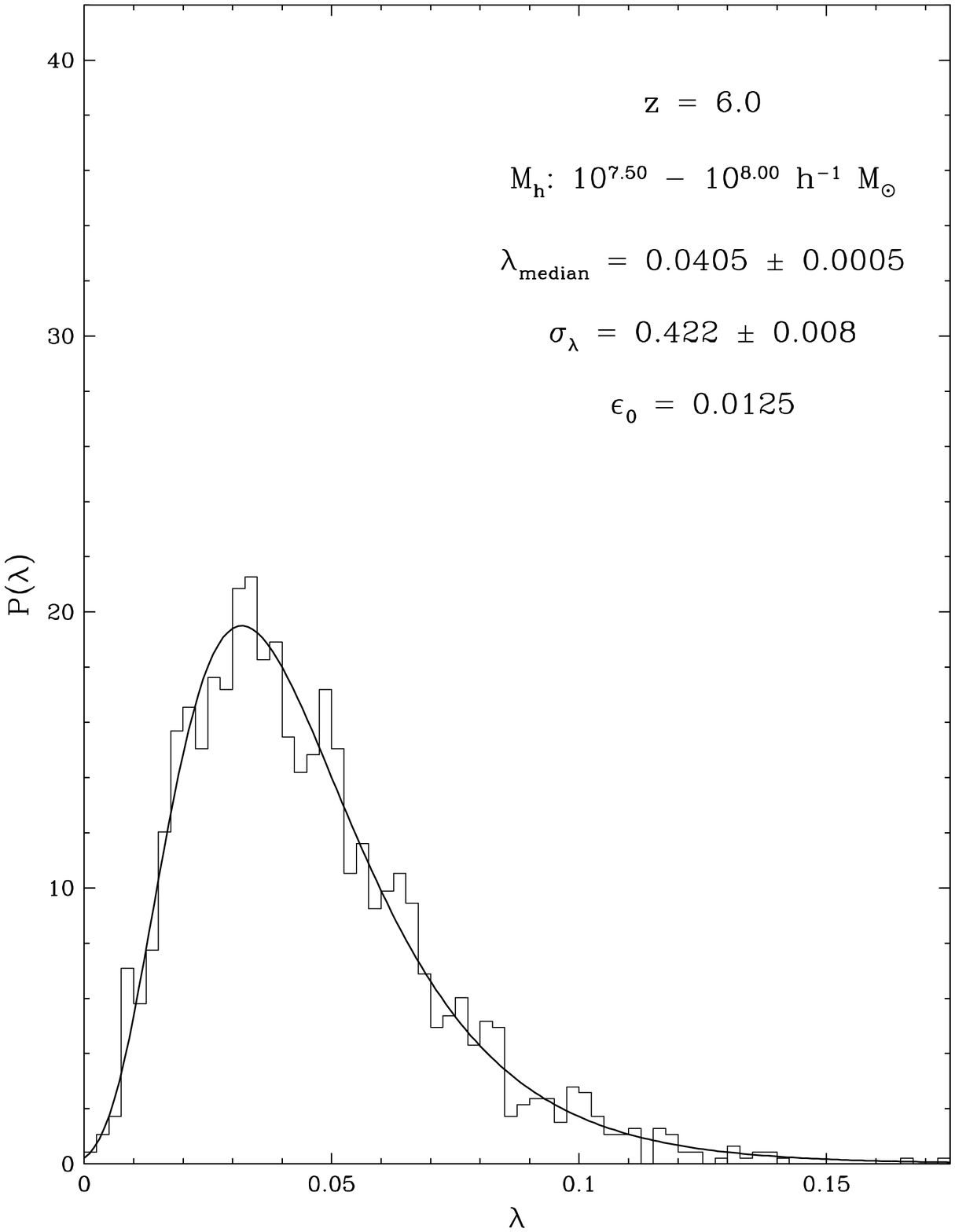}
\vskip 0.01in
\includegraphics[width=2.8in]{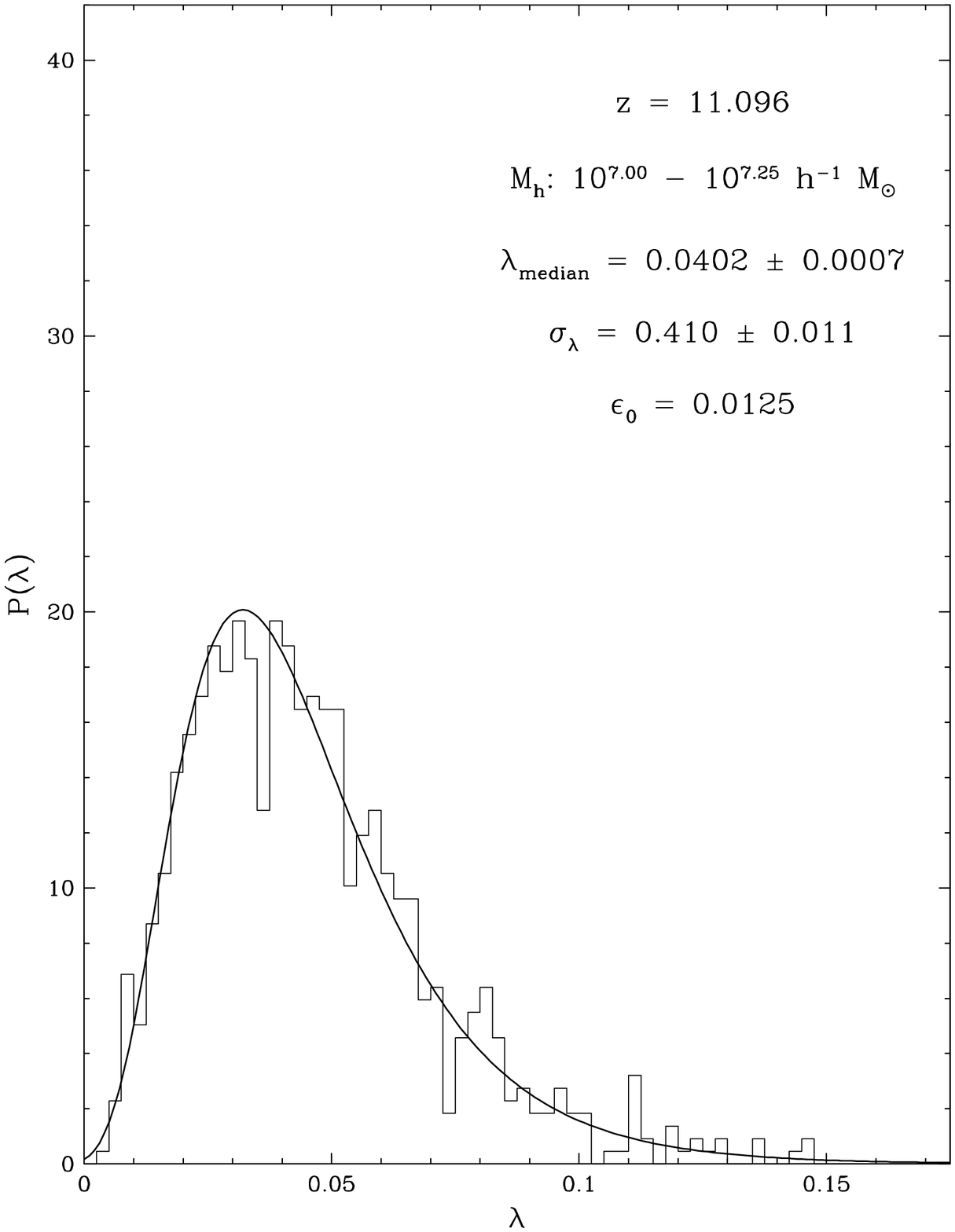}
\hspace{0.01in}
\includegraphics[width=2.8in]{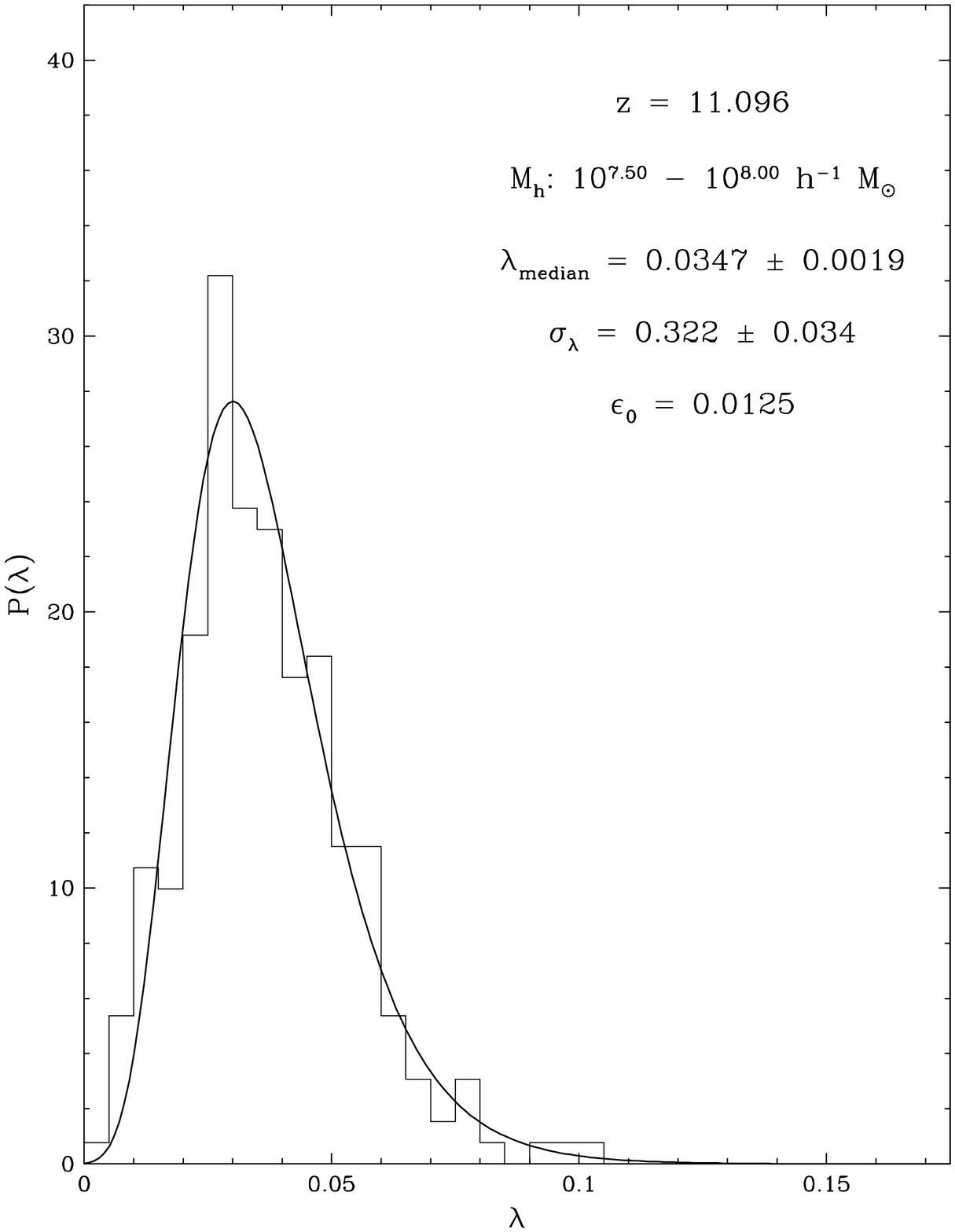}
\caption{The distributions of halo spin parameter $\lambda$ 
for four randomly selected cases upon different mass and redshift. The modified lognormal fits 
are shown as smooth curves.
\label{f9}}
\end{figure*}

\epsscale{1.0}
\begin{figure}
\includegraphics[width=1.0\hsize,angle=270]{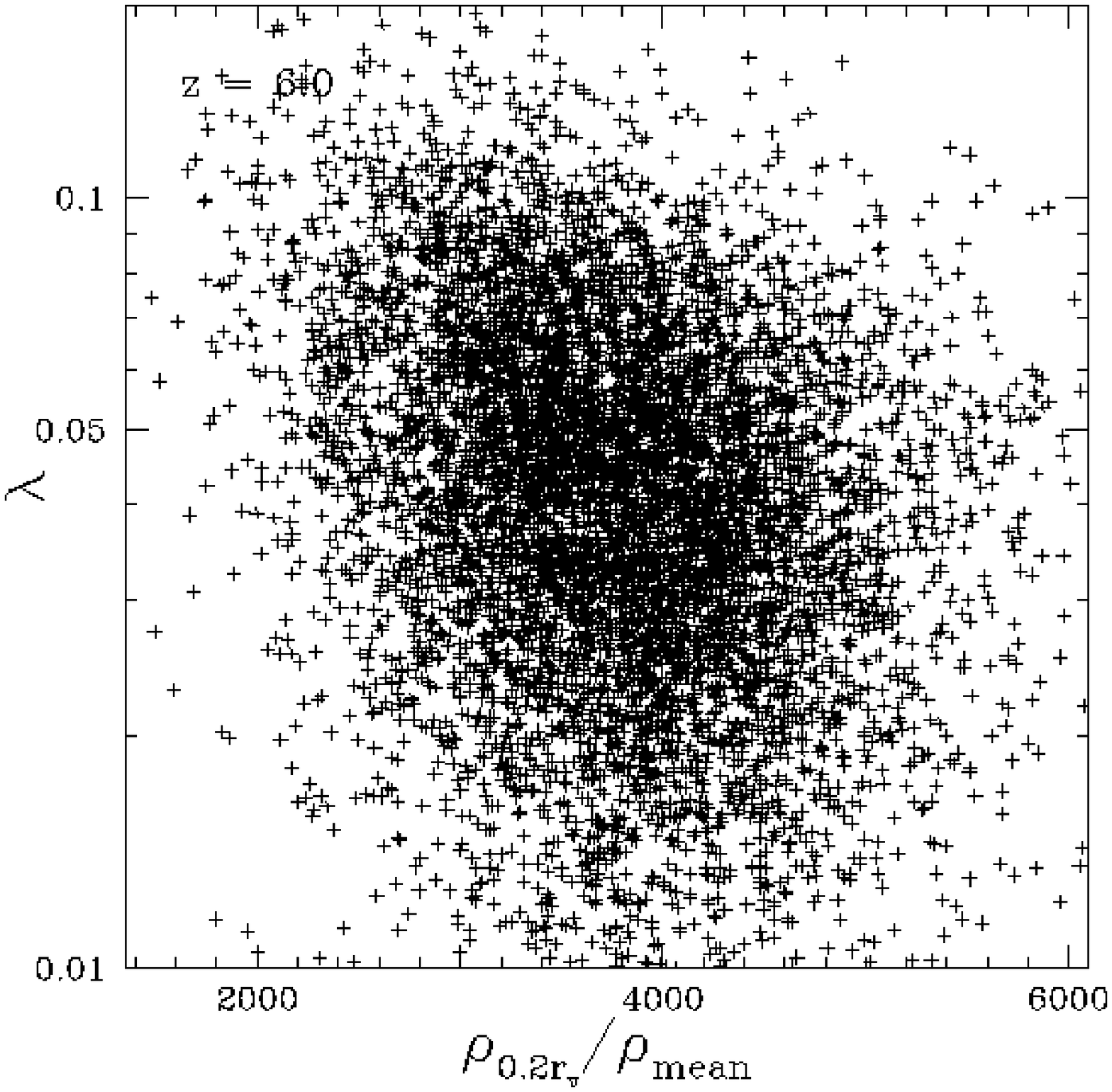}
\caption{
A scatter plot of the spin parameter, $\lambda$,
versus the halo central density, defined as the density at $r<0.2r_{v}$. 
for all the halos with $M>10^6 \ h^{-1} M_\odot$ at $z=6$. 
\label{f10b}}
\end{figure}

\epsscale{1.0}
\begin{figure}
\plotone{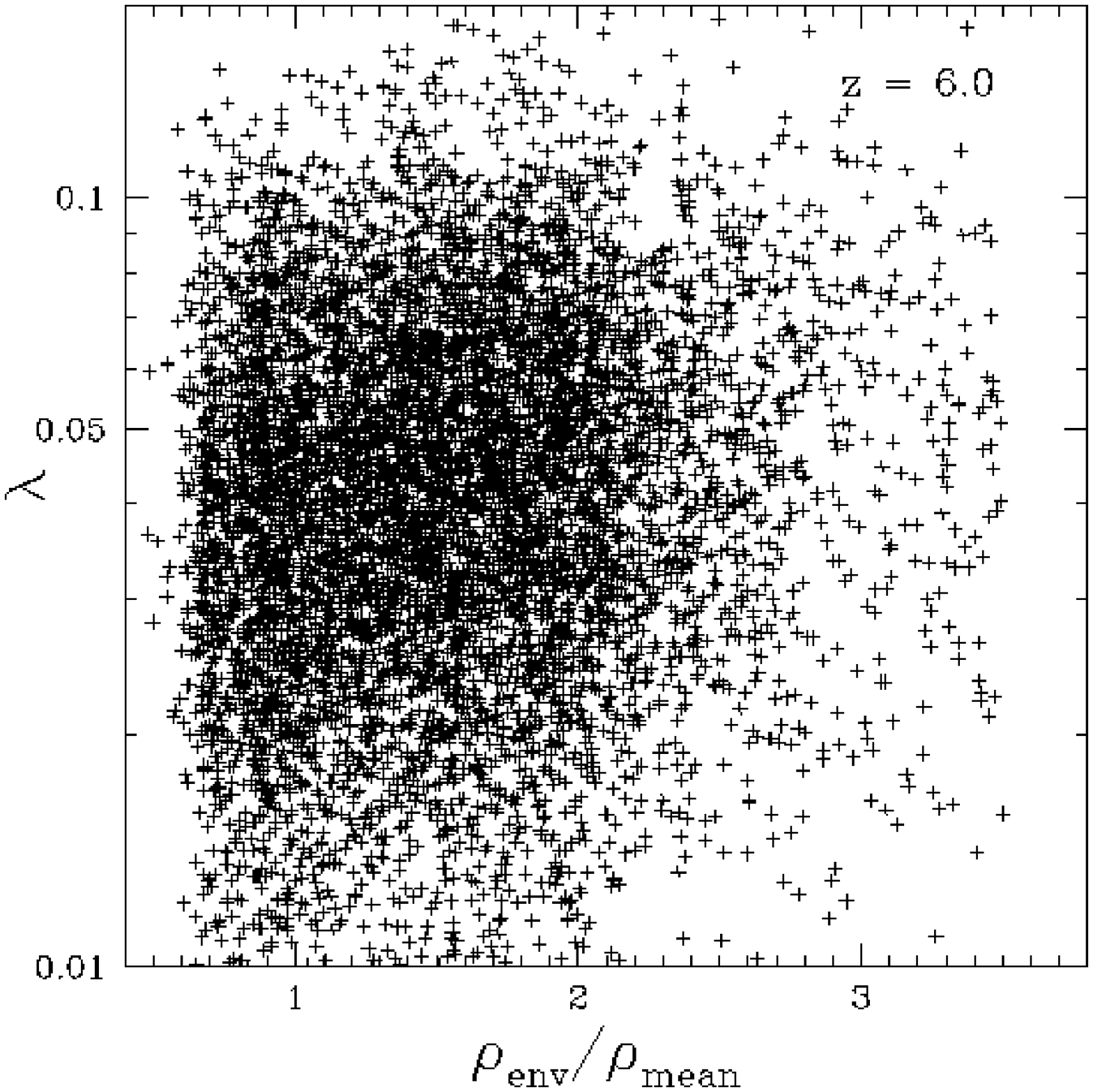}
\caption{
A scatter plot of the spin parameter, $\lambda$,
versus the environmental density, defined as the dark matter density
smoothed by a gaussian window of radius $0.3h^{-1}$Mpc,
for all the halos with $M>10^6 \ h^{-1} M_\odot$ at $z=6$. 
\label{f10b}}
\end{figure}

\epsscale{1.0}
\begin{figure}
\includegraphics[width=1.0\hsize,angle=270]{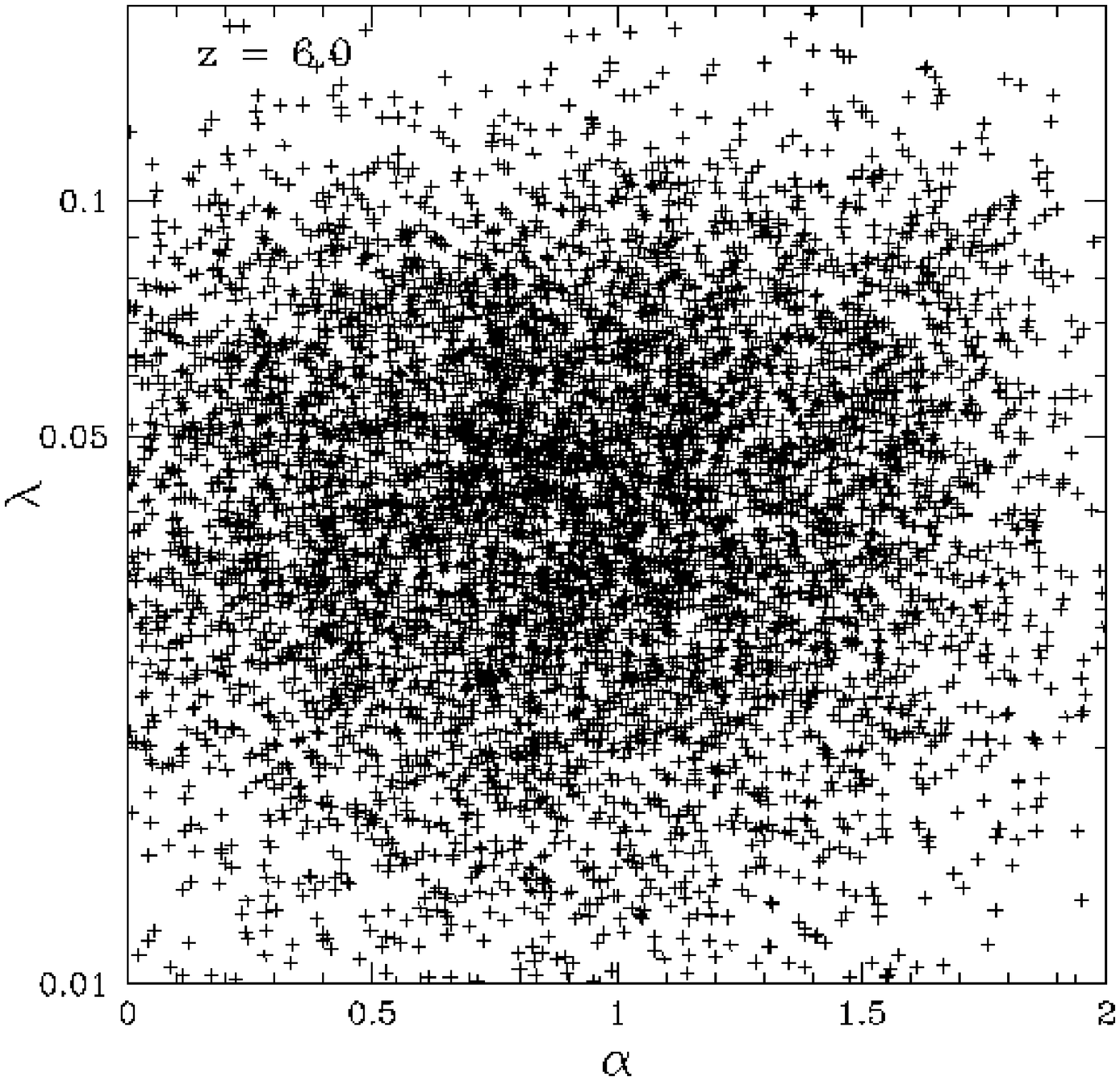}
\caption{Comparison of the halo density profile slope parameter $\alpha$ 
versus spin parameter $\lambda$. No correlation between the two is observed.
\label{f11}}
\end{figure}

\begin{deluxetable}{rrrrrrrr} 
\tablecolumns{5} 
\tablewidth{0pc} 
\tablecaption{Spin Parameter : $\lambda_{median}$(M,z)} 
\tablehead{ 
\colhead{Halo Mass ($h^{-1}$ M$_\odot$) } & \colhead{\ \ \ z=6.0 \ \ }   & \colhead{\ \ z=7.4 \ }  & 
\colhead{z=9.08 \ } & \colhead{\ \ z=11.096}}
\startdata 
10$^{7.00}$ - 10$^{7.25}$ \ \ \ \ \ & \ 0.043$\pm$0.001 & \ \ \ 0.043$\pm$0.001 \ \ & \ \ 0.042$\pm$0.001 \ \ & 
\ 0.040$\pm$0.001 \\ 
10$^{7.25}$ - 10$^{7.50}$ \ \ \ \ \ & \ 0.042$\pm$0.001 & \ \ \ 0.041$\pm$0.001 \ \ & \ \ 0.037$\pm$0.001 \ \ & 
\ 0.037$\pm$0.001 \\ 
10$^{7.50}$ - 10$^{8.00}$ \ \ \ \ \ & \ 0.041$\pm$0.001 & \ \ \ 0.039$\pm$0.001 \ \ & \ \ 0.035$\pm$0.001 \ \ & 
\ 0.035$\pm$0.002 \\ 
$>$ 10$^{8.00}$ \ \ \ \ \ \ \ \ \  & \ 0.035$\pm$0.001 & \ \ \ 0.033$\pm$0.001 \ \ & \ \ 0.031$\pm$0.002 \ \ & 
\ 0.031$\pm$0.007 \\ 
\hline
\enddata 
\end{deluxetable} 

\begin{deluxetable}{rrrrrrrr} 
\tablecolumns{5} 
\tablewidth{0pc} 
\tablecaption{Spin Parameter : $\sigma_{\lambda}$(M,z)} 
\tablehead{ 
\colhead{Halo Mass ($h^{-1}$ M$_\odot$)} & \colhead{\ \ \ z=6.0 \ \ }   & \colhead{\ \ z=7.4 \ }  & 
\colhead{z=9.08 \ \ } & \colhead{\ z=11.096}}
\startdata 
10$^{7.00}$ - 10$^{7.25}$ \ \ \ \ \ & \ 0.43$\pm$0.01 & \ \ \ 0.44$\pm$0.01 \ \ & \ \ 0.44$\pm$0.01 \ \ & 
\ 0.41$\pm$0.01 \\ 
10$^{7.25}$ - 10$^{7.50}$ \ \ \ \ \ & \ 0.43$\pm$0.01 & \ \ \ 0.42$\pm$0.01 \ \ & \ \ 0.38$\pm$0.01 \ \ & 
\ 0.41$\pm$0.02 \\ 
10$^{7.50}$ - 10$^{8.00}$ \ \ \ \ \ & \ 0.42$\pm$0.01 & \ \ \ 0.41$\pm$0.01 \ \ & \ \ 0.41$\pm$0.01 \ \ & 
\ 0.32$\pm$0.03 \\ 
$>$ 10$^{8.00}$ \ \ \ \ \ \ \ \ \  & \ 0.40$\pm$0.01 & \ \ \ 0.39$\pm$0.02 \ \ & \ \ 0.31$\pm$0.04 \ \ & 
\ 0.43$\pm$0.15 \\ 
\hline
\enddata 
\end{deluxetable} 

\begin{deluxetable}{rrrrrrrr} 
\tablecolumns{5} 
\tablewidth{0pc} 
\tablecaption{Angular Momentum Profile : $A_0$(M,z)} 
\tablehead{ 
\colhead{Halo Mass ($h^{-1}$ M$_\odot$)} & \colhead{\ \ \ z=6.0 \ \ }   & \colhead{\ \ z=7.4 \ }  & \colhead{z=9.08 \ \ } & \colhead{\ z=11.096}}
\startdata 
10$^{7.0}$ - 10$^{7.5}$ \ \ \ \ \ & \ 0.51$\pm$0.03 & \ \ \ 0.49$\pm$0.04 \ \ & \ \ 0.44$\pm$0.06 \ \ & 
\ 0.44$\pm$0.08 \\ 
10$^{7.5}$ - 10$^{8.0}$ \ \ \ \ \ & \ 0.55$\pm$0.06 & \ \ \ 0.52$\pm$0.12 \ \ & \ \ 0.48$\pm$0.14 \ \ & 
\ 0.50$\pm$0.16 \\ 
$>$ 10$^{8.0}$ \ \ \ \ \ \ \ \ \ & \ 0.56$\pm$0.09 & \ \ \ 0.51$\pm$0.21 \ \ & \ \ 0.49$\pm$0.24 \ \ & 
\ 0.45$\pm$0.32 \\ 
\hline
\enddata 
\end{deluxetable} 

\begin{deluxetable}{rrrrrrrr} 
\tablecolumns{5} 
\tablewidth{0pc} 
\tablecaption{Angular Momentum Profile : $\sigma_{A}$(M,z)} 
\tablehead{ 
\colhead{Halo Mass ($h^{-1}$ M$_\odot$) } & \colhead{\ \ \ z=6.0 \ \ }   & \colhead{\ \ z=7.4 \ }  & \colhead{z=9.08 \ \ } & \colhead{\ z=11.096}}
\startdata 
10$^{7.0}$ - 10$^{7.5}$ \ \ \ \ \ & \ 0.27$\pm$0.03 & \ \ \ 0.28$\pm$0.04 \ \ & \ \ 0.27$\pm$0.06 \ \ & 
\ 0.25$\pm$0.07 \\ 
10$^{7.5}$ - 10$^{8.0}$ \ \ \ \ \ & \ 0.26$\pm$0.05 & \ \ \ 0.29$\pm$0.10 \ \ & \ \ 0.29$\pm$0.11 \ \ & 
\ 0.26$\pm$0.12 \\ 
$>$ 10$^{8.0}$ \ \ \ \ \ \ \ \ \ & \ 0.27$\pm$0.08 & \ \ \ 0.35$\pm$0.18 \ \ & \ \ 0.29$\pm$0.16 \ \ & 
\ 0.36$\pm$0.23 \\ 
\hline
\enddata 
\end{deluxetable} 

We compute the spin parameter defined as
\begin{equation}
\lambda \equiv {J|E|^{1/2}\over G M^{5/2}}
\end{equation}
\noindent
(Peebles 1969),
where $G$ is the gravitational constant;
$M$ is the total mass of the dark matter halo;
$J$ is the total angular momentum of the dark matter halo;
$E$ is the total energy of the dark matter halo;
all quantities are computed within the virial radius.
We fit the $\lambda$ distributions using a modified
lognormal function:
\begin{equation}
P(\lambda) = {1\over (\lambda+\epsilon_0)\sqrt{2\pi} \sigma_{\lambda}}\exp{(-{[\ln (\lambda+\epsilon_0) -\ln \lambda_0]^2\over 2\sigma_{\lambda}})},
\end{equation}
\noindent
where $\epsilon_0$ is fixed to be $0.0125$,
which is determined through experimentation.
The modified lognormal fits for $\lambda$ are shown
as smooth curves in Figure 15;
the goodness of the fits is typical.
Tables 8,9 lists fitting parameters 
$\lambda_0$ and $\sigma_\lambda$, respectively. 
Note that the median value of $\lambda$
is $\lambda_{med}=\lambda_0-\epsilon_0$.

We see that the typical spin parameter has a value
$0.03-0.04$.
However, the distribution of the spin parameter among
halos is very broad, with a lognormal dispersion of $\sim 0.4$.
This implies that consequences that depend on 
the spin of a halo are likely to be widely distributed
even at a fixed dark matter halo mass.
Such consequences may include the size of a galactic disk 
and correlations between dark matter halo spin
(conceivable misalignment between spin of gas and spin
of dark matter would complicate the situation)
and other quantities.

In Figures 16 and 17 we show the correlation
between $\lambda$ and the central density of the halo,
and between $\lambda$ and the environmental density, respectively.
We find that there may possibly exist 
a weak correlations between 
$\lambda$ and the central density of the halo,
in the sense that halos with higher central densities (or equivalently
earlier formation times) have lower $\lambda$,
with a very large scatter,
whereas no correlation is discernible between $\lambda$
and the environmental density.
Finally, Figure 18 show the relation between $\lambda$ and $\alpha$,
where no correlation is visible.

\begin{figure*}
\includegraphics[width=2.8in]{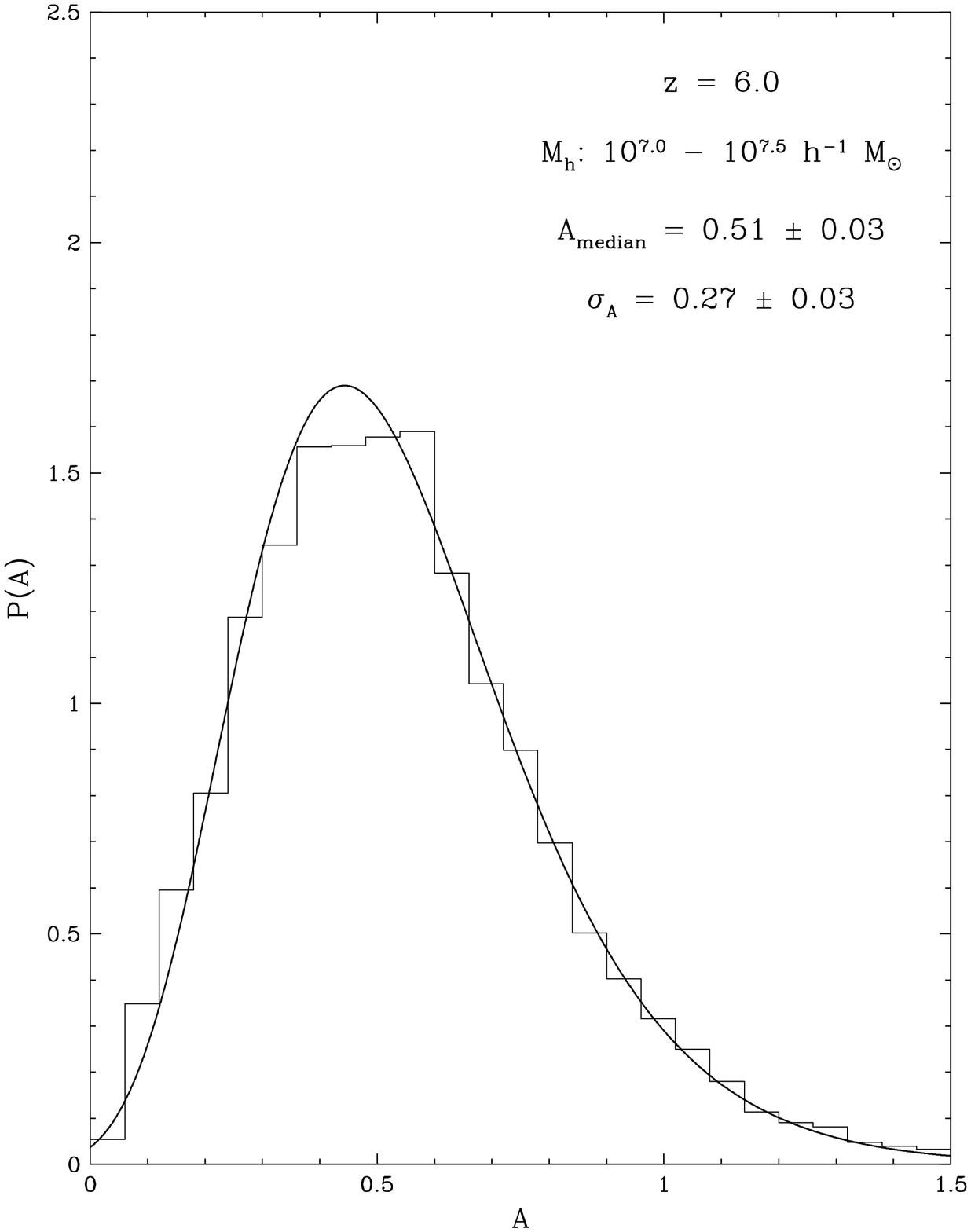}
\hspace{0.01in}
\includegraphics[width=2.8in]{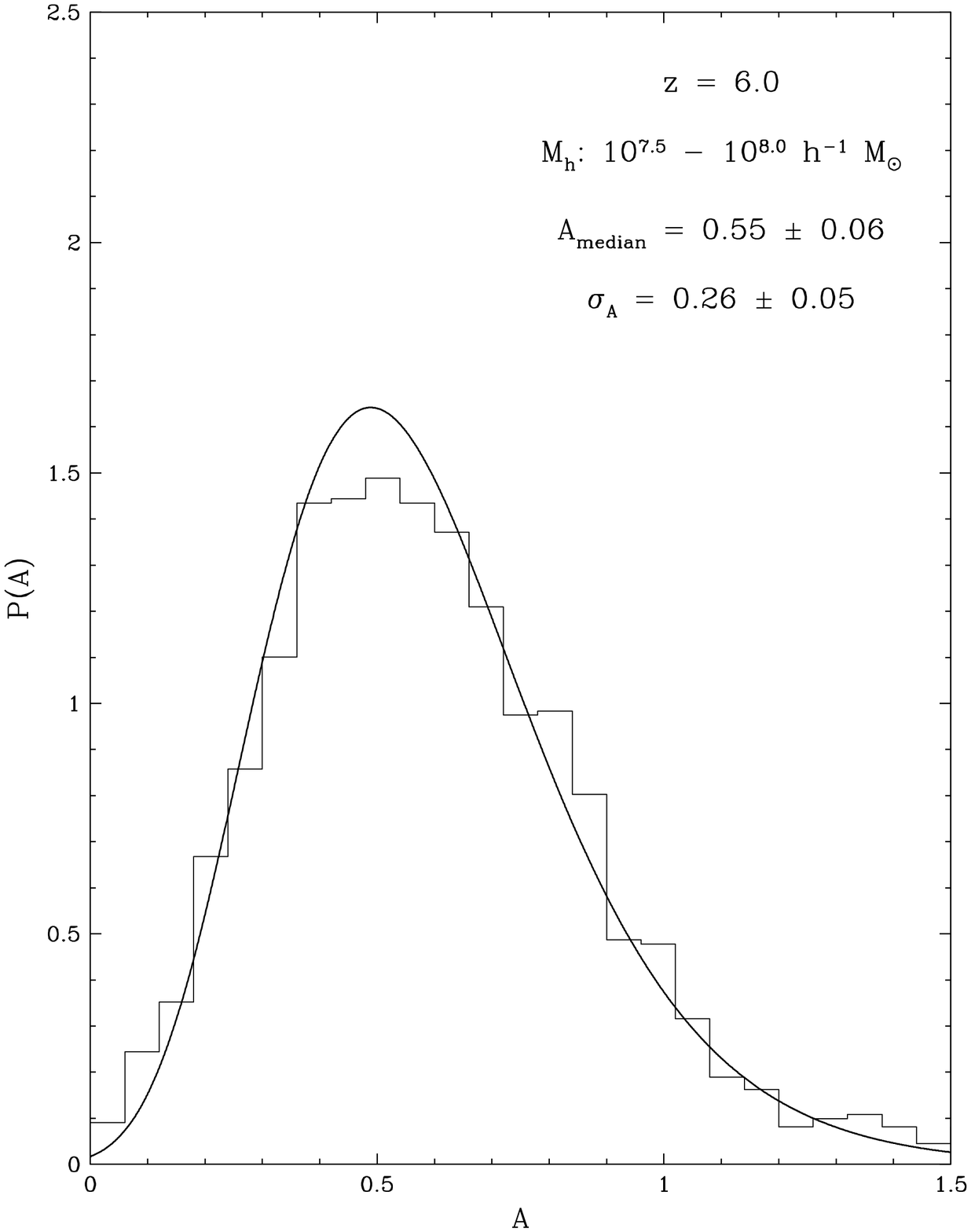}
\vskip 0.01in
\includegraphics[width=2.8in]{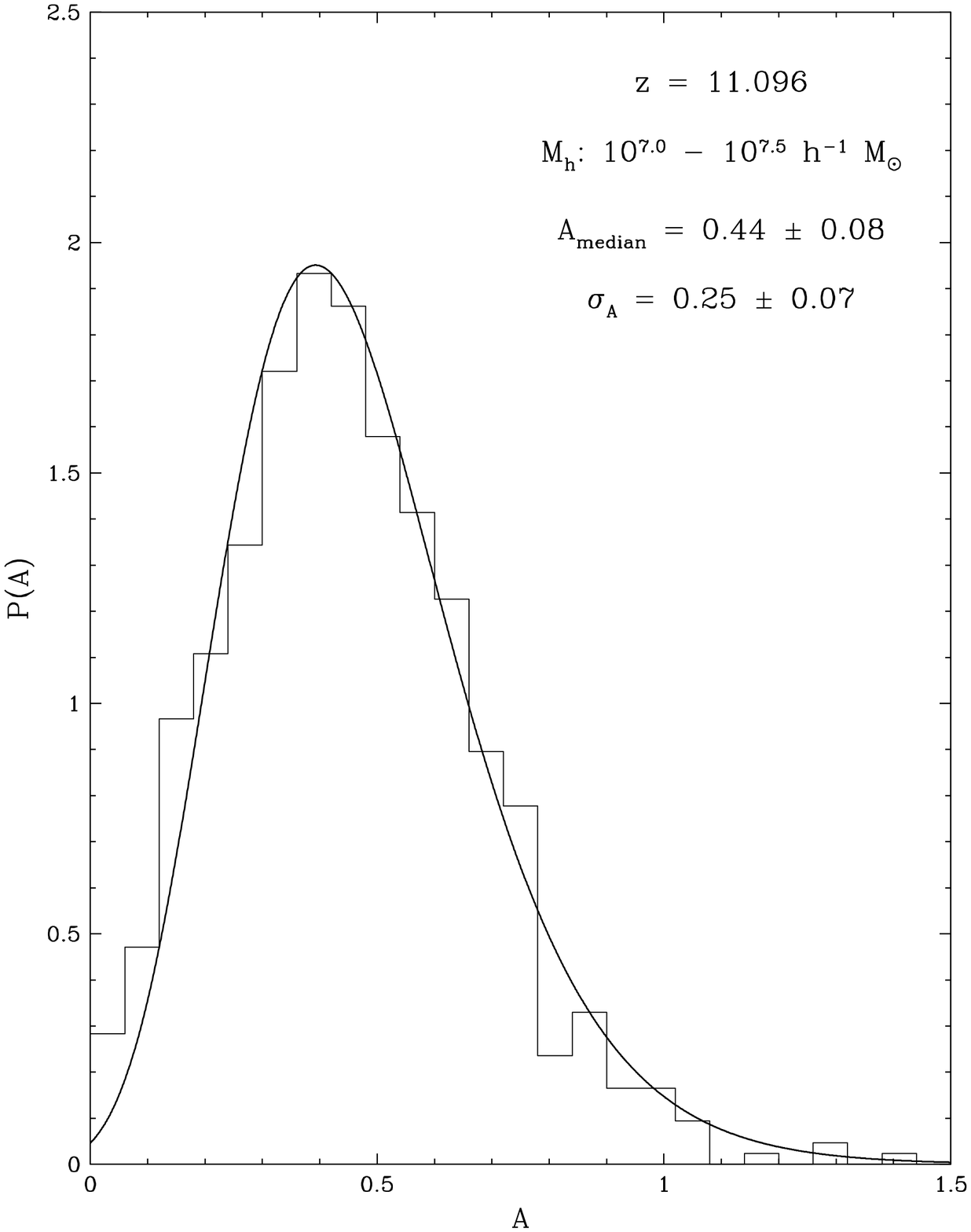}
\hspace{0.01in}
\includegraphics[width=2.8in]{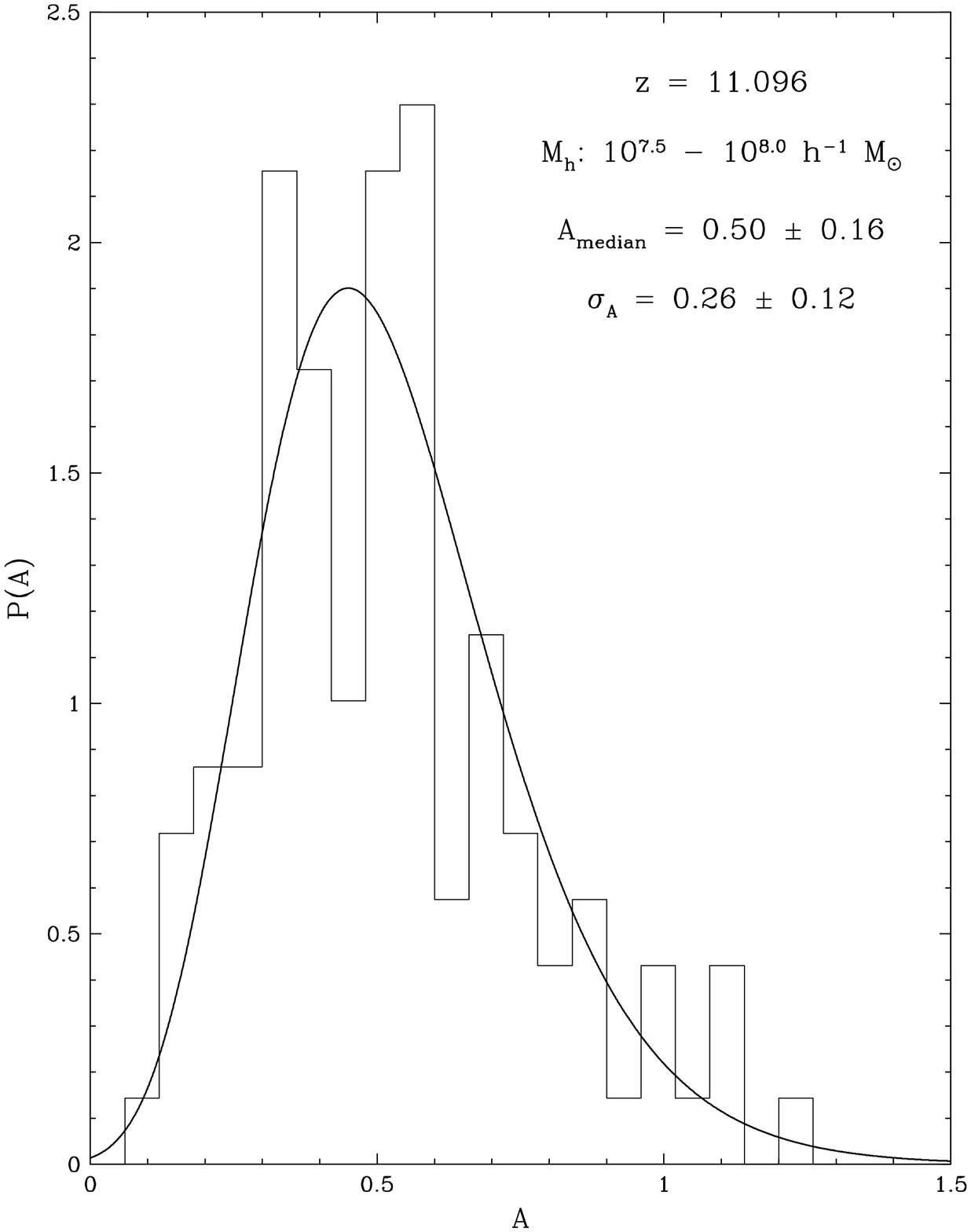}
\caption{The distributions of slope parameter $A$ in the halo angular momentum profile fitting 
for four randomly selected cases upon different mass and redshift. The lognormal fits 
are shown as smooth curves.
\label{f12}}
\end{figure*}

\begin{figure*}
\includegraphics[width=2.8in]{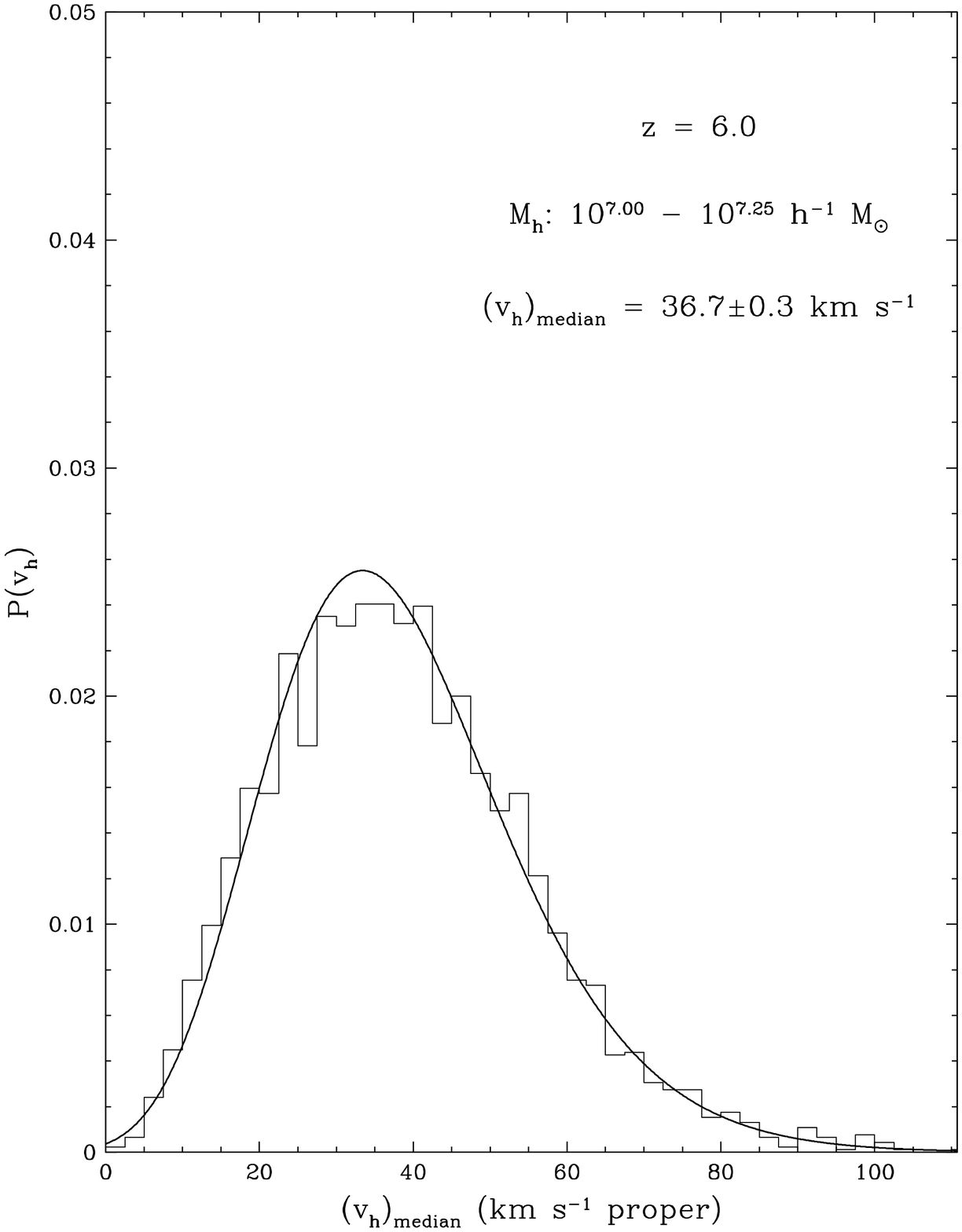}
\hspace{0.01in}
\includegraphics[width=2.8in]{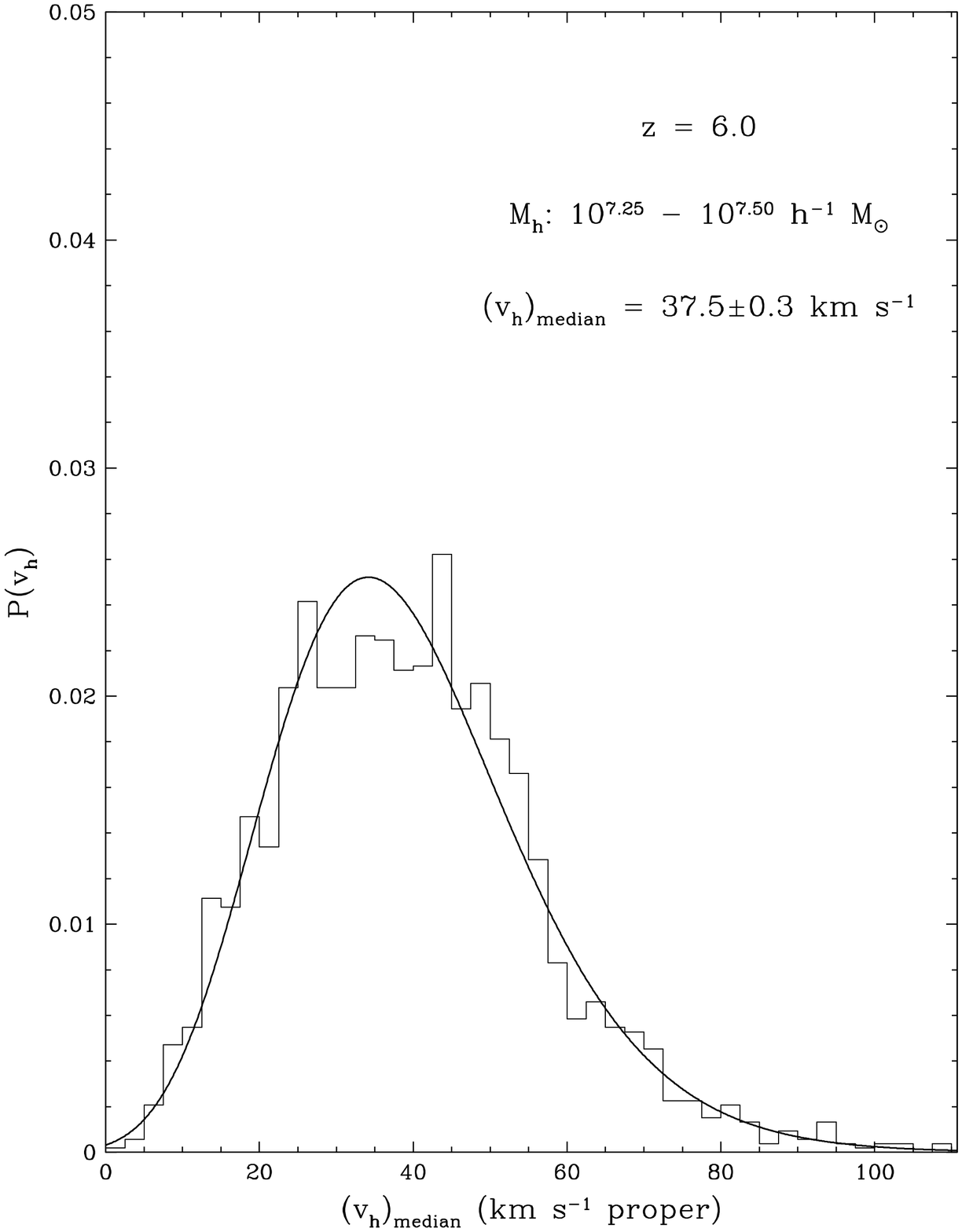}
\vskip 0.01in
\includegraphics[width=2.8in]{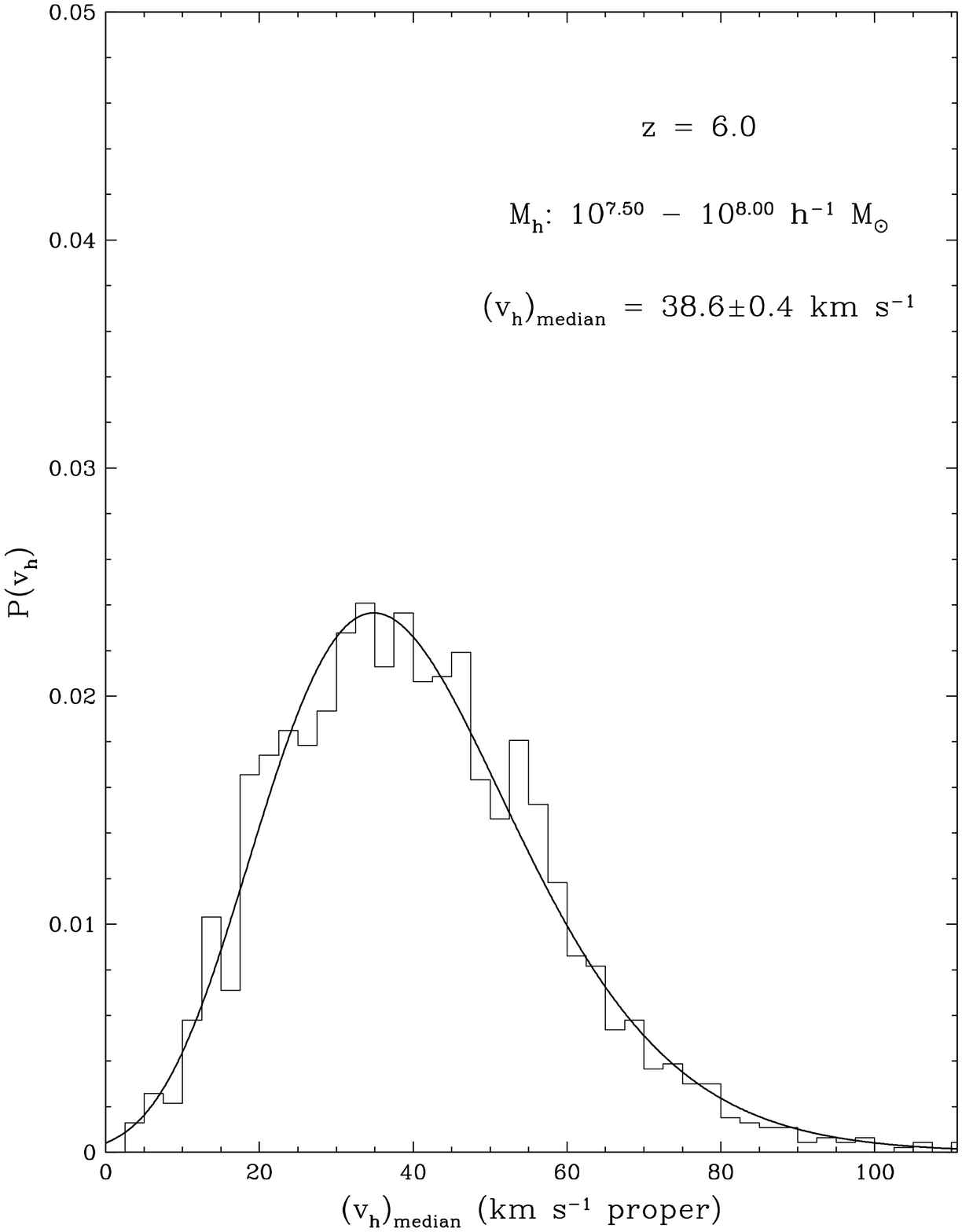}
\hspace{0.01in}
\includegraphics[width=2.8in]{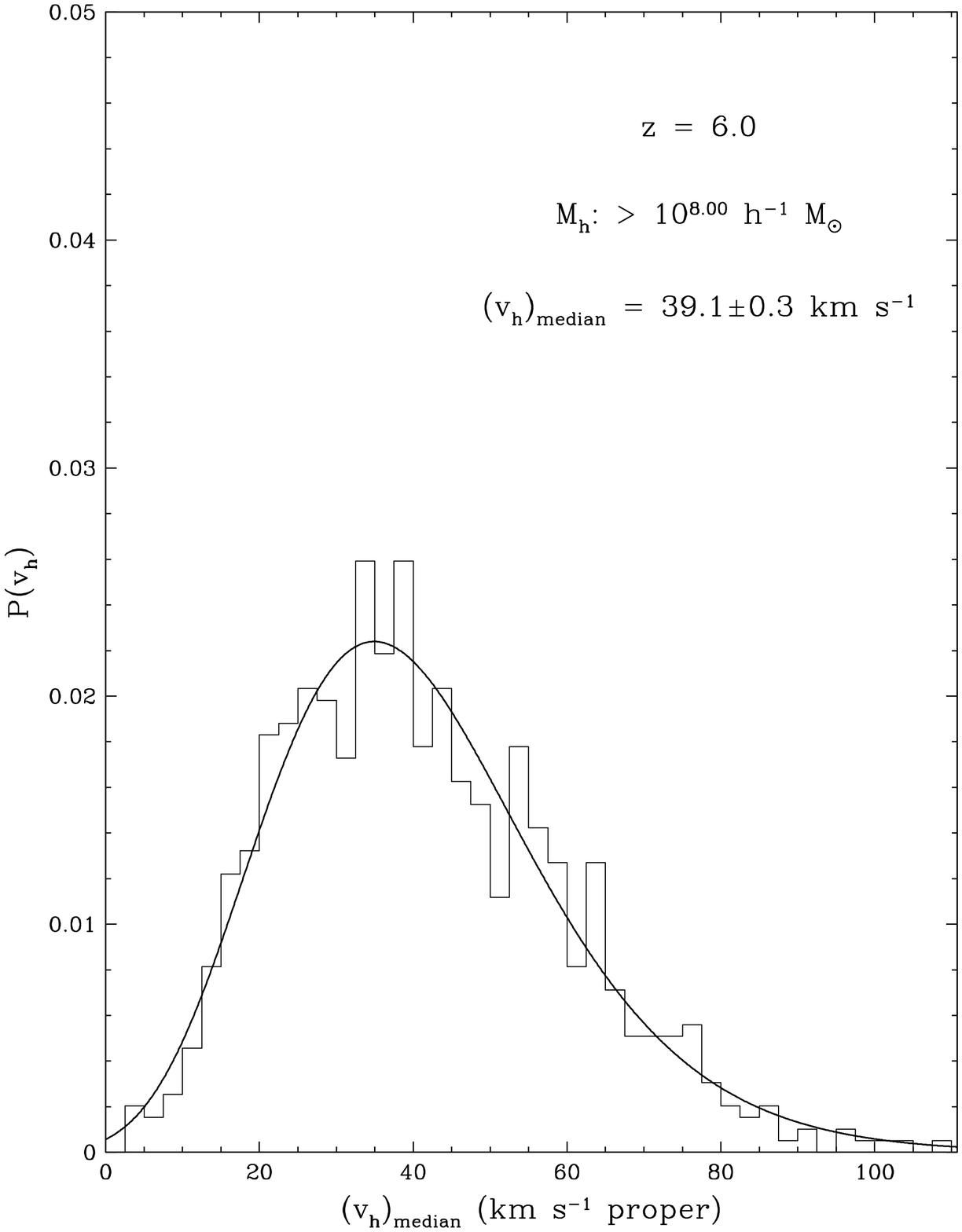}
\caption{
The distributions of dark matter halo peculiar velocity 
for four mass bins at $z=6$.
The lognormal fits are shown as smooth curves.
\label{f12}}
\end{figure*}
\subsection{Angular Momentum Profile For Dark Matter Halos}

Next, we compute the angular momentum profiles
for individual dark matter halos.
We then fit the each angular momentum profile by the following function in the 
small $|j|$ regime:
\begin{equation}
{M(<j)\over M_v} = A{j\over j_0} + {M(<0)\over M_v},
\end{equation}
\noindent
where $A$ and $M(<0)$ are two fitting parameters;
$M_v$ is the virial mass and
$j_0\equiv J/M_v=\lambda GM^{3/2}/|E|^{1/2}$.

In order to compute the angular momentum profile
an appropriate smoothing window needs to be
applied to dark matter particles.
We find that $M(<0)/M_v$ 
varies for different smoothing scales, with typical values around 0.2,  
while $A$ remains roughly constant for each individual halo.
However, $A$ is broadly distributed for all dark matter halos.
Figure 19 show histograms for the distributions of $A_0$,
for four randomly selected cases.
We fit the $A$ distribution using a modified
lognormal function:
\begin{equation}
P(A) = {1\over (A+\epsilon_A)\sqrt{2\pi} \sigma_{A}}\exp{(-{[\ln (A+\epsilon_A) -\ln A_0]^2\over 2\sigma_{A}})},
\end{equation}
\noindent
where $\epsilon_A=0.4$ is fixed through experimentation.
Tables 10,11 list fitting parameters 
$A_0$ and $\sigma_A$, respectively. 

In general we find our fitting formula (Equation 10)
provides a good fit for each individual halo.
The distribution of matter at small $j$ is most relevant for
the formation of central objects, such as black holes
(e.g., Colgate \etal 2003) or bulges (e.g., D'Onghia \& Burkert 2004).
Our calculation indicates that
the fraction of mass in a halo having specific angular momentum
less than a certain value is
roughly $0.5$ times the ratio of that value over the average
specific angular momentum of the halo.

\subsection{Bulk Velocity of Dark Matter Halos}

Finally, we compute the peculiar velocity of dark matter halos.
We find that the distribution once again can be fitted by
lognormal distributions as equation (9).
In order to provide
a good fit for the results at $z=6$
shown in Figure 20, 
it is found that $\epsilon=40$km/s in equation (8)
with median velocity $v_m\sim 38\pm 2$km/s and 
lognormal dispersion $\sigma_v=0.22\pm 0.01$.
We caution, however, the absolute value of the peculiar
velocity of each halo, unlike the quantities examined in 
previous subsections,
may be significantly affected by large waves not present in
our simulation box.
Adding missing large waves should increase the zero point $\epsilon$
to a larger value.
Therefore, the peculiar velocity shown should be
treated as a lower limit.
In other words, expected peculiar velocity 
of dark matter halos at these redshifts
are likely in excess of $30-40$km/s.


\section{Conclusions}

Using a high resolution TPM N-body simulation of the
standard cold dark matter cosmological model
with a particle mass of $m_p=3.57\times 10^4 \ h^{-1}M_\odot$
and a softening length of $\epsilon=0.14 \ h^{-1}$kpc
in a $4h^{-1}$Mpc box,
we compute various properties of dark matter halos 
with mass $10^{6.5}-10^9\msun$ at redshift $z=6-11$.
We find the following results.

\noindent (1) Dark matter halos at such small mass 
at high redshifts are already significantly biased
over matter with a bias factor in the range $2-6$.

\noindent (2) The dark matter halo mass function displays a 
slope at the small end $2.05\pm 0.15$.

\noindent (3) The central density profile of dark matter halos
are found to be in the range $(0.4-1.0)$
well fitted by 
$\alpha_0=0.75((1+z)/7.0)^{-1.25}(M/10^7\msun)^{0.11(1+z)/7.0}$
with a dispersion of $\pm 0.5$,
in rough agreement with the theoretical
arguments given in Ricotti (2003) and Subramanian \etal (2000).

\noindent (4) The median spin parameter of the dark matter halos
is $0.03-0.04$ but with a lognormal dispersion of $\sim 0.4$.
The angular momentum profile at the small end
is approximately linear
with the fraction of mass in a halo having specific angular momentum
less than a certain value is
roughly $0.5$ times the ratio of that value over the average
specific angular momentum of the halo.

\noindent (5) The dark matter halos move at a typical velocity in excess of $30-40$km/s.

\acknowledgments
This research was supported in part by AST-0206299 and NAG5-13381.
The computations were performed on the National Science Foundation
Terascale Computing System at the Pittsburgh Supercomputing Center.


\clearpage

\end{document}